\documentclass[11pt]{article}
\usepackage{amsfonts,amsmath,amssymb}
\usepackage{enumerate}
\usepackage[hidelinks]{hyperref}
\usepackage{bbm}
\usepackage{nicefrac}
\usepackage[all]{xy}
\usepackage{graphicx}
\usepackage{bm}
\usepackage{makecell}
\usepackage[table]{xcolor}

\usepackage{upgreek}
\usepackage{slashed}

\usepackage{float}			% NEW Helps position tables right here [H]
\usepackage{lscape}                 % NEW creates pages in landscape format

\usepackage{booktabs}

\newcommand{\be}{\begin{equation}}
\newcommand{\ee}{\end{equation}}

\newcommand{\bea}{\begin{eqnarray}}
\newcommand{\eea}{\end{eqnarray}}

\newcommand{\bes}{\begin{subequations}}
\newcommand{\ees}{\end{subequations}}

\newcommand{\cN}{{\cal N}}

\def\ft#1#2{{\textstyle{{\scriptstyle #1}\over {\scriptstyle #2}}}}

\def\sst#1{{\scriptscriptstyle #1}}

\def\0{{\sst{(0)}}}
\def\1{{\sst{(1)}}}
\def\2{{\sst{(2)}}}
\def\3{{\sst{(3)}}}
\def\4{{\sst{(4)}}}
\def\5{{\sst{(5)}}}
\def\6{{\sst{(6)}}}
\def\7{{\sst{(7)}}}
\def\8{{\sst{(8)}}}

\def\cV{{{\cal V}}}
\def\cM{{{\cal M}}}

\allowdisplaybreaks  %Introduces page breaks inside \eqnarray%

\usepackage{multirow}
\usepackage{rotating}

\newcommand{\ba}{\begin{align}}
\newcommand{\ea}{\end{align}}

\newcommand{\bse}{\begin{subequations}}
\newcommand{\ese}{\end{subequations}}

\allowdisplaybreaks

\usepackage{lipsum}

\newlength\Colsep
\begin{document}

\makeatletter
\renewcommand{\theequation}{\thesection.\arabic{equation}}
\@addtoreset{equation}{section}
\makeatother

\begin{titlepage}

\begin{flushright}
%
% Preprint no.
%
%\today
\end{flushright}

\vspace{5pt}

   \begin{center}
   \baselineskip=16pt

   \begin{Large}\textbf{
\hspace{-18pt} Trombone gaugings of five-dimensional maximal supergravity %\\[8pt]
}
   \end{Large}

\vspace{25pt}

{\large  Oscar Varela }

\vspace{30pt}

	\begin{small}

   {\it Department of Physics, Utah State University, Logan, UT 84322, USA}

	\vspace{5pt}
          
   {\it Instituto de F\'\i sica Te\'orica UAM/CSIC, 28049 Madrid, Spain} 
		
	\end{small}

\vskip 70pt

\end{center}

\begin{center}
\textbf{Abstract}
\end{center}

\begin{quote}

The maximal gauged supergravities in five spacetime dimensions with gauge groups contained in $\mathbb{R}^+ \times \textrm{E}_{6(6)}$ are described. The $\textrm{E}_{6(6)}$ factor is the duality group of ungauged maximal supergravity and $\mathbb{R}^+$ is the scaling symmetry of the five-dimensional metric, usually called the \textit{trombone} symmetry. The equations of motion and supersymmetry variations for these supergravities are given to lowest order in fermions, and the mass matrices are provided. Then, the theories with gauge groups contained in a maximal subgroup of $\mathbb{R}^+ \times \textrm{E}_{6(6)}$ are classified, and a new family of such supergravities uncovered. For a concrete theory in this class, some supersymmetric anti-de Sitter vacua are found and their mass spectra computed within the gauged supergravity. These vacua are argued to be related to superconformal phases of the M5-brane field theory.

\end{quote}

\vfill

\end{titlepage}

\tableofcontents

%%%%%%%%%%%%%%%%%%%%%%%%%%%%%%%%%%%%%%%%%%%%%%%%%%%%%%%%%%%%%%%%%%%%%%%%%%

%%%%%%%%%%%%%%%
%%%%%%%%%%%%%%%

\section{Introduction}

%%%%%%%%%%%%%%%
%%%%%%%%%%%%%%%

Supergravity theories in various dimensions (see \cite{Sezgin:2023hkc} for a recent survey) provide natural venues to model a wide range of phenomena in model building, cosmology or holography. In five spacetime dimensions, maximal ungauged supergravity was constructed long ago \cite{Cremmer:1979uq}. The Lagrangian displays an $\mathrm{E}_{6(6)}$ exceptional symmetry, which is enhaced to $\mathbb{R}^+ \times \mathrm{E}_{6(6)}$ at the level of the field equations. The $\mathbb{R}^+$ factor is a symmetry of the equations of motion, but not of the Lagrangian, under scalings of the metric and all the other supergravity fields with appropriate weights. This $\mathbb{R}^+$ scaling symmetry is enjoyed by all ungauged supergravity theories, and has been popularly known as {\it trombone} symmetry following \cite{Cremmer:1997xj}.

The $\mathbb{R}^+ \times \mathrm{E}_{6(6)}$ duality of ungauged $D=5$ $\cN=8$ supergravity acts on the field equations as a global symmetry. As any other supergravity, this theory can be gauged, namely, a subgroup $G \subset \mathbb{R}^+ \times \mathrm{E}_{6(6)}$ may be promoted to act locally in a way consistent with supersymmetry. For any supergravity, the gauging procedure entails turning on additional interaction terms, including non-abelian field strengths, fermion masses, Yukawa couplings or scalar self-interactions,
 see \cite{Trigiante:2016mnt} for a review. For five-dimensional maximal supergravity specifically, the general gaugings have been detailed in \cite{deWit:2004nw}. This reference recovered some previously known gauged supergravities, particularly those with gauge groups SO$(p,q)$  \cite{Gunaydin:1984qu,Gunaydin:1985cu,Pernici:1985ju} and In\"on\"u-Wigner contractions CSO$(p,q,r)$ thereof \cite{Andrianopoli:2000fi}, and developed the formalism to deal with the most general gauging of $D=5$ $\cN=8$ supergravity.

While \cite{deWit:2004nw} laid down the general formalism, that reference mainly dealt with gaugings contained only in $\mathrm{E}_{6(6)}$. The general features of gaugings involving also the trombone symmetry $\mathbb{R}^+$ have been discussed in \cite{LeDiffon:2008sh}, but a more explicit account of gaugings of $D=5$ $\cN=8$ supergravity with gauge group contained in the full duality group $\mathbb{R}^+ \times \mathrm{E}_{6(6)}$ of the field equations is still lacking. In contrast, in three \cite{LeDiffon:2008sh} and four \cite{LeDiffon:2011wt} dimensions, the maximal gauged supergravities involving the local scaling symmetry $\mathbb{R}^+$ have been completely specified building on the general formalisms of \cite{Nicolai:2000sc} and \cite{deWit:2007mt}. More recently, the trombone gaugings of four-dimensional half-maximal supergravity have been characterised \cite{Liatsos:2024ehz} building on the general formalism of \cite{DallAgata:2023ahj} (see also \cite{Schon:2006kz}).

The purpose of this paper is to fill this gap by specifying the most general $D=5$ $\cN=8$ gauged supergravity with gauge group contained in both factors $\mathbb{R}^+ \times \mathrm{E}_{6(6)}$ of the ungauged duality group. A discussion of the trombone gaugings of maximal five-dimensional supergravity is timely in the light of \cite{Bhattacharya:2024tjw}, which recently encountered a particular theory in this class by dimensional reduction of $D=11$ supergravity. The presentation of general aspects will rely on the formalism already developed to great detail in \cite{deWit:2004nw,LeDiffon:2008sh} and, for that reason, will be brief. The central new result of section \ref{sec:GenGauge} is the presentation of the trombone contributions to the $D=5$ $\cN=8$ field equations, supersymmetry transformations, and mass matrices of the various supergravity fields. 

The remainder of the paper contains further new material. Section \ref{sec:TCSO} addresses the most general gauging strictly contained in the $\mathbb{R}^+ \times \mathrm{SL}(2,\mathbb{R}) \times \mathrm{SL}(6,\mathbb{R}) $ maximal subgroup of $\mathbb{R}^+ \times \mathrm{E}_{6(6)}$. In doing so, a novel family of gaugings, dubbed TCSO$(p,q,r;\rho)$ for non-negative integers $p$, $q$, $r$, $\rho$, is uncovered. This extends the tromboneless CSO$(p,q,r)$ gaugings \cite{Gunaydin:1984qu,Gunaydin:1985cu,Pernici:1985ju,Andrianopoli:2000fi} to the trombone case, and encompasses the gauging of \cite{Bhattacharya:2024tjw} as the TCSO$(5,0,1;1)$ representative in this class. The latter gauging is fixed in section \ref{eq:AdSvac}, where a convenient subsector is described that contains interesting supersymmetric anti-de Sitter (AdS) vacua. These include the half-maximally supersymmetric, $\cN=4$, vacuum already discussed in \cite{Bhattacharya:2024tjw}, that uplifts to the $D=11$ AdS$_5$ solution with sixteen supercharges found by Maldacena and N\'u\~nez (MN2) in \cite{Maldacena:2000mw} in relation with the M5-brane field theory. It also includes an $\cN=2$ AdS vacuum that uplifts to the eight-supercharge $D=11$ AdS$_5$ solution (MN1) also found in \cite{Maldacena:2000mw}. More generally, the model contains a one-parameter family of $\cN=2$ vacua that uplifts to the eight-supercharge family of $D=11$ AdS$_5$ solutions obtained by Bah, Beem, Bobev and Wecht (BBBW) \cite{Bah:2011vv,Bah:2012dg}, with hyperbolic Riemann surface therein. Finally, the mass spectrum of these solutions within five-dimensional maximal gauged supergravity is computed, using the mass matrices of section \ref{sec:GenGauge}.

%%%%%%%%%%%%%%%
%%%%%%%%%%%%%%%

\section{General $D=5$ $\cN=8$ gauged supergravity} \label{sec:GenGauge}

%%%%%%%%%%%%%%%
%%%%%%%%%%%%%%%

The ingredients that go into five-dimensional maximal supergravity are reviewed in section \ref{sec:5DFields} in order to set up conventions. Section \ref{sec:Eoms} presents the field equations and supersymmetry variations to lowest order in fermions. Finally, the vacuum conditions and mass matrices are derived in section \ref{sec:MassMat}. The presentation follows the $D=5$ $\cN=8$ notations and conventions of \cite{deWit:2004nw,LeDiffon:2008sh}, with occasional minor typographical changes and the gauge coupling there set to $g=1$.

%%%%%%%%%%%%%%%

\subsection{Field content and embedding tensor} \label{sec:5DFields}

%%%%%%%%%%%%%%%

The field content of maximal supergravity in five spacetime dimensions includes bosons and fermions of all spin between 0 and 2. The former comprise the metric, $g_{\mu \nu} = \eta_{a b} \, e_\mu{}^a e_\nu{}^b$, with vielbein $e_\mu{}^a$, as well as 27 gauge fields $A^M$ and 42 scalars sitting in a coset representative $\cV_M{}^{ij}= \cV_M{}^{[[ij]]}$ of $\mathrm{E}_{6(6)}/\mathrm{USp}(8)$. The fermion sector contains 8 gravitini, $\psi_\mu{}^i$, and 48 spin-$1/2$ fermions, $\chi^{ijk}=\chi^{[[ijk]]}$, where spacetime spinor indices have been supressed. Double brackets are used to denote symplectic antisymmetrisation, $\chi^{[[ijk]]} \,  \Omega_{ij}=0$, where $\Omega_{ij}$ is the $\mathrm{USp}(8)$-invariant form. The latter allows $\mathrm{USp}(8)$ indices to be lowered and raised. One may additionally consider higher-rank tensor fields including 27 two-form potentials $B_{M}$ along with three-, four- and five-form potentials that make up a so-called `tensor hierarchy' \cite{deWit:2008ta}. These are typically Hodge-dual to lower-rank tensors \cite{Bergshoeff:2009ph} and, except for the two-forms which play a crucial role in the gauged theory \cite{deWit:2004nw}, they will be subsequently ignored. The indices $\mu = 0 , 1  , \ldots , 4$ and $a = 0 , 1 , \ldots ,4$ are local and global $D=5$ spacetime indices, while the internal indices $M = 1 , \ldots, 27$ and $i = 1 , \ldots , 8$ respectively label the (anti)fundamental representations of the duality group E$_{6(6)}$ and its maximal compact subgroup USp$(8)$. 

Table \ref{tab:5DSummary} summarises this field content. The bosonic and fermionic fields respectively lie in representations of E$_{6(6)}$ and USp$(8)$. In addition, they transform under non-compact $\mathbb{R}^+$ rescalings with the charges indicated as subscripts in the table. The ungauged $D=5$ $\cN=8$ supergravity is described by a Lagrangian that features all of the above fields except the two-forms, which may be excluded. This Lagrangian is only E$_{6(6)}$-invariant, as it scales under $\mathbb{R}^+$ transformations. The scaling is homogeneous, though, as the entire Lagrangian scales with definite $\mathbb{R}^+$ charge. As a consequence, the equations of motion are invariant under the full group of transformations $\mathbb{R}^+ \times \mathrm{E}_{6(6)}$. By construction, the Bianchi identities satisfied by the gauge and tensor hierarchy fields are also $\mathbb{R}^+ \times \mathrm{E}_{6(6)}$-invariant. Thus, $\mathbb{R}^+ \times \mathrm{E}_{6(6)}$ is a symmetry of the combined set of field equations, including equations of motion and Bianchi identities, of $D=5$ $\cN=8$ ungauged supergravity.

The group $\mathbb{R}^+ \times \mathrm{E}_{6(6)}$ acts as a global symmetry in the ungauged theory. Gauged supergravities are obtained by promoting a subgroup thereof to act locally. New couplings and interactions must be turned on  in this gauging process in a way compatible with maximal supersymmetry. These  include non-abelian gauge interactions, Yukawa and mass terms, and scalar self-interactions. All these couplings are governed by the components, $\Theta_{M}{}^\alpha$ and $\vartheta_M$, of the embedding tensor \cite{deWit:2004nw,LeDiffon:2008sh},
\begin{equation} \label{eq:XSymbolsGen}
X_M^{\bm{r}_w} = -w \, \vartheta_M \, t_0 + \big( \Theta_{M}{}^\alpha +  \tfrac92 (t^\alpha)_{M}{}^{Q} \,  \vartheta_{Q} \big)  t_\alpha \; .
\end{equation}
The latter is a constant object that couples the gauge fields $A^M$ to the $\mathbb{R}^+ \times \mathrm{E}_{6(6)}$ generators $t_0$, $t_\alpha$, $\alpha= 1 , \ldots , 78$, in the gauging-induced covariant derivatives $D=d-A^M X_M^{\bm{r}_w} $. In (\ref{eq:XSymbolsGen}), the generators need to be evaluated in the  representations $\bm{r}_w$ of $\mathbb{R}^+ \times \mathrm{E}_{6(6)}$, summarised in table~\ref{tab:5DSummary}, appropriate to the fields on which the covariant derivative acts. For example, acting on the vielbein, the gravitino, or the scalar coset, one has
\begin{equation} \label{eq:VariousX}
X_M^{\bm{1}_1} = - \, \vartheta_M \; , \qquad
X_M^{\bm{1}_{\frac12}} = -\tfrac12 \, \vartheta_M \; , \qquad
(X_M^{\bm{27}_0})_N{}^P = \Theta_{M}{}^\alpha \, (t_\alpha)_{M}{}^{N} +  \tfrac92 \,   \mathbb{P}^Q{}_M{}^P{}_N \, \vartheta_Q \; , 
\end{equation}
while acting on gauge fields,
\begin{eqnarray} \label{eq:defX}
X_{M N}{}^{P} =
\Theta_{M N}{}^{P}
 +  \Big(  \tfrac92 \,  \mathbb{P}^Q{}_M{}^P{}_N  - \delta_{M}^{Q}\delta_{N}^{P} \Big)
\vartheta_{Q}  \; , \quad 
\textrm{with \, $\Theta_{M N}{}^{P} \equiv \Theta_M{}^\alpha (t_\alpha)_N{}^P$,}
\end{eqnarray}
where the representation label has been dropped, as costumary. In these expressions, $(t_\alpha)_M{}^N$ are the $\mathrm{E}_{6(6)}$ generators in the fundamental representation (see appendix \ref{sec:E6Conventions} for conventions) and the projector to the adjoint representation, $\mathbb{P}^K{}_M{}^L{}_N \equiv (t^\alpha)_M{}^{K} (t_\alpha)_N{}^{L}$, has been introduced. Note for later use that this projector satisfies the identity 
\begin{equation} \label{eq:ProjId}
\mathbb{P}^K{}_M{}^L{}_N =
\ft1{18}\,\delta_M^{K}\delta_N^{L} +\ft1{6}\,\delta_M^{L}\delta_N^{K} 
-\ft53\, d_{MNP}\,d^{KLP}
\; ,
\end{equation}
in terms of the E$_{6(6)}$ cubic invariants $d_{MNP}$ and $d^{MNP}$. Finally, indices in the adjoint of $\mathrm{E}_{6(6)}$ are raised and lowered with the Killing-Cartan form, $\kappa_{\alpha \beta} \equiv \mathrm{tr} ( t_{\alpha}t_{\beta} )$. Most of the discussion below will hinge around the embedding tensor (\ref{eq:defX}) acting on gauge fields. 

\begin{table}[]

\centering

%\resizebox{\textwidth}{!}{

\begin{tabular}{l | cccccccc}
%
%\hline
%\\[-2.5mm]
%
 	& $g_{\mu\nu} $ &  $\psi_\mu^i $ &  $A_\mu^M$ & $B_{\mu\nu M}$ & $\chi^{ijk}$ & $\cV_M{}^{ij}$ & $\Theta_{M}{}^\alpha$ & $\vartheta_M$ 
\\[2pt]
\hline
\\[-10pt]
$\mathbb{R}^+ \times \mathrm{E}_{6(6)}$ &	$\bm{1}_2$  &  $\bm{1}_{\frac12} $   &	$\overline{\bm{27}}_1$ & $\bm{27}_2$  & $\bm{1}_{-\frac12}$  & $\bm{27}_0$  & $\overline{\bm{351}}_{-1}$  & $\bm{27}_{-1}$  
\\[10pt]
$\mathbb{R}^+ \times \mathrm{USp}(8)$ &	$\bm{1}_2$  &  $\overline{\bm{8}}_{\frac12} $   &	$\bm{1}_1$ & $\bm{1}_2$  & $\overline{\bm{48}}_{-\frac12}$  & $\overline{\bm{27}}_0$ & $\bm{1}_{-1}$ & $\bm{1}_{-1}$ 
\\[10pt]
%
%
%\hline
%
\end{tabular}

%}

\caption{\footnotesize{Charges under $\mathbb{R}^+ \times \mathrm{E}_{6(6)}$ and $\mathbb{R}^+ \times \mathrm{USp}(8)$ of the supergravity fields and embedding tensor.
}\normalsize}
\label{tab:5DSummary}
\end{table}

The embedding tensor components $\Theta_{M}{}^\alpha$ and $\vartheta_M$ must satisfy linear and quadratic constraints \cite{deWit:2004nw,LeDiffon:2008sh}. The former are required by compatibility with maximal supersymmetry, and restrict them to lie in the $\overline{\bm{351}}_{-1}$ and the $\bm{27}_{-1}$ representations of $\mathbb{R}^+ \times \mathrm{E}_{6(6)}$, respectively. For completeness, these charges are also shown in table \ref{tab:5DSummary}. While $\vartheta_M$ naturally sits in the $\bm{27}_{-1}$ as the notation already indicates, $\Theta_{M}{}^\alpha$ formally lies in the $( \bm{27} \times \bm{78} )_{-1} = \bm{27}_{-1} + \overline{\bm{351}}_{-1} + \bm{1728}_{-1}$ representation. The linear constraint thus enforces $P_{\overline{\bm{351}} \, M}{}^{\alpha N}{}_\beta \,   \Theta_{N}{}^\beta = \Theta_{M}{}^\alpha $, where the projector $P_{\overline{\bm{351}} \, M}{}^{\alpha N}{}_\beta$ to the $\overline{\bm{351}}$ of $\mathrm{E}_{6(6)}$ can be found in \cite{deWit:2002vt}. Equivalently, the $\overline{\bm{351}}_{-1}$ components of the embedding tensor can be  encoded in an antisymmetric matrix $Z^{MN} = Z^{[MN]} $ related to $\Theta_{M}{}^\alpha$ by either expression \cite{deWit:2004nw,LeDiffon:2008sh}
\begin{equation} \label{eq:ZembTen}
Z^{MN} = \Theta_P{}^\alpha \, (t_\alpha)_Q{}^M \, d^{NPQ}  \; , \qquad 
Z^{MN} = 2 \, \Theta_P{}^\alpha \, (t_\alpha)_Q{}^T \, d^{PRM} \, d^{QSN} \, d_{RST} \; .
\end{equation}
Usage of either $\Theta_{M}{}^\alpha$ or $Z^{MN}$ will alternate below based on convenience.

The quadratic constraints must be imposed in turn so that the embedding tensor (\ref{eq:defX}) acting on gauge fields closes into a Lie algebra,
\begin{equation} \label{eq:LieAlg}
[ X_{M} , X_{N}] = - X_{M N}{}^{P} \, X_{P} \; ,
\end{equation}
namely, the algebra of the physical gauge group $G$. These quadratic constraints read \cite{deWit:2004nw,LeDiffon:2008sh}
\begin{eqnarray} \label{eq:QCs}
& \overline{\bm{27}}_{-2} & : \;  Z^{MN} \vartheta_N +\tfrac{15}{2}  d^{MNP} \vartheta_N \vartheta_P = 0  \; , \nonumber \\[3pt]
& \bm{351}_{-2} & : \;  Z^{PQ} \vartheta_R  \, d_{MPK} \, d_{NQL} \,  d^{KLR} =0  \; , \\[3pt]
& \bm{\overline{27}}_{-2} + \bm{\overline{1728}}_{-2} & : \;  4(t_\alpha)_K{}^L\,Z^{KR}Z^{NS}d_{RSL}
+3(t_\alpha)_K{}^L\,Z^{KN}\vartheta_L
+3(t_\alpha)_K{}^N\,Z^{KL}\vartheta_L = 0 \; ,  \nonumber 
\end{eqnarray}
and sit in the indicated $\mathbb{R}^+ \times \mathrm{E}_{6(6)}$ representations. 

A few comments on field redundancies and duality frames are in order to conclude this section. Firstly, the embedding tensor formalism used here contains redundant fields in the vector-tensor sector. This redundancy can be fixed at the expense of loosing manifest E$_{6(6)}$ convariance. In doing so, any $D=5$ $\cN=8$ gauged supergravity can accordingly be formulated in terms of only $\mathrm{rank} \, X_{MN}{}^P$ gauge fields and $27-\mathrm{rank} \, X_{MN}{}^P$ two-forms, with $X_{MN}{}^P$ the embedding tensor (\ref{eq:defX}) acting on gauge fields. See~\cite{deWit:2004nw} for further details. Secondly, the symmetry is no longer $\mathbb{R}^+ \times \mathrm{E}_{6(6)}$ in the gauged supergravity, as it gets reduced instead to the local gauge group $G \subset \mathbb{R}^+ \times \mathrm{E}_{6(6)}$. In common parlance, the gauging $G$ is activated from the ungauged theory by first choosing a duality frame and then turning on the interactions dictated by the embedding tensor. This is just language to indicate that two gauged supergravities whose embedding tensors $X_{MN}{}^P$ and $\tilde{X}_{MN}{}^P$ are tensorially related by an $\mathrm{E}_{6(6)}$ transformation $U_M{}^N$ through
\begin{equation} \label{eq:TransXSymbol}
\tilde{X}_{MN}{}^P = U_M{}^Q \, U_N{}^R \, X_{QR}{}^S  \, (U^{-1})_S{}^P \; , 
\end{equation}
are equivalent. Couplings may look different depending on either embedding tensor used to formulate the theory, but the physics stays the same. Sections \ref{sec:TCSO} and \ref{eq:AdSvac} will discuss aspects of the same gauged supergravity formulated in different duality frames.

%%%%%%%%%%%%%%%

\subsection{Field equations and supersymmetry transformations} \label{sec:Eoms}

%%%%%%%%%%%%%%%

It follows from (\ref{eq:defX}) that supergravities with $\vartheta_M = 0$ have gauge groups $G$ strictly contained in E$_{6(6)}$. As reviewed above, the latter is a symmetry of the ungauged Lagrangian and, therefore, these gauged supergravities admit a Lagrangian as well. These gaugings were the focus of \cite{deWit:2004nw}, where they were analysed in detail and the Lagrangian and supersymmetry transformations were presented. In contrast, supergravities with $\vartheta_M \neq 0$ necessarily have the $\mathbb{R}^+$ trombone symmetry gauged, along with a subgroup of E$_{6(6)}$. For this reason, these gauged supergravities do not admit a Lagrangian and must be described at the level of the field equations. 

The determination of the physical couplings induced by the gauging follows an algorithmic process that has been extensively specified in the literature, see~{\it e.g.}~\cite{Trigiante:2016mnt} for a review. These couplings arise in the supersymmetry variations of the fermions and in the equations of motion of both fermions and bosons (or, equivalently if $\vartheta_M = 0$, in the Lagrangian). The gauging-induced terms in the supersymmetry variations can be fixed by imposing the supersymmetry algebra, namely, by forcing the commutator of two supersymmetries to close into combinations of bosonic symmetries. The gauging-induced couplings in the equations of motion can in turn be determined by ensuring that the fermionic and bosonic field equations transform into each other under supersymmetry. The specific task of turning on trombone contributions to gauged supergravities is greatly simplified if the couplings induced by gaugings strictly contained in E$_{d(d)}$ were previously known. In this case, only the new $\vartheta_M$ contributions to the supersymmetry variations and field equations need to be tracked down following the process summarised above. 

This strategy was carefully explained in \cite{LeDiffon:2008sh,LeDiffon:2011wt} and, for that reason, the discussion here will be brief. I will simply present the field equations and supersymmetry transformations for general $D=5$ $\cN=8$ gauged supergravity including trombone gaugings that are obtained by turning on $\vartheta_M$ in their $\vartheta_M=0$ counterparts of \cite{deWit:2004nw}. Some details on their derivation can be found in appendix \ref{sec:susyEoms}. Reassuringly, the bosonic equations of motion found by this process agree with those obtained by other methods in \cite{Blair:2024ofc}. Without further ado, the bosonic equations of motion (Einstein, gauge field, tensor and scalar) read, to lowest order in fermions,
{\setlength\arraycolsep{0pt}
\begin{eqnarray}
\label{eq:EinsteinEOM} && R_{\mu \nu} = - \tfrac{1}{24} D_\mu M_{MN} D_\nu M^{MN}  + \tfrac{1}{3} g_{\mu\nu} V + \tfrac{1}{2 }  M_{MN} \big(  H_{\mu \rho}{}^M H_{\nu}{}^\rho{}^N  - \tfrac{1}{6} g_{\mu\nu}  \, H_{\rho \sigma}{}^M \, H^{\rho \sigma N}  \big)  \; ,  \\[12pt]
\label{eq:vectorEOM} && D \big( M_{MN} *H_\2^N \big) + \tfrac{1}{6} \,\big( \Theta_{MN}{}^P - \tfrac{27}{2} \delta_M^P \vartheta_N  \big)  \, M^{NQ} * D M_{PQ} + \tfrac{\sqrt{5}}{4} \, d_{MNP} H_\2^N \wedge H_\2^P =0 \,  ,  \\[12pt]
\label{eq:tensorEOM} && Z^{MN} \big( H_{\3 N} + \sqrt{\tfrac{2}{5}} \, M_{NP} * H_\2^P \big) =0  \; ,  \\[12pt]
\label{eq:scalarEOM} && P_{MN}{}^{PQ} \hspace{-1pt} \Big( \hspace{-2pt}
D*DM_{PQ}  \hspace{-2pt}  - \hspace{-2pt}  M^{RS} \hspace{-2pt}  DM_{PR} \hspace{-1pt}  \wedge \hspace{-2pt}  *DM_{QS}  \hspace{-2pt} - \hspace{-2pt} 12 \, M_{PR} M_{QS} \, H_\2^R \wedge * H_\2^S \hspace{-2pt}  + \hspace{-2pt} 12 \, \tilde{V}_{PQ} \, \textrm{vol}
\Big) =0 . \nonumber 
\\
\end{eqnarray}
}These must be supplemented by the Bianchi identities
\begin{equation}
D H_\2^{M} = \big(  Z^{MN} - \tfrac{15}{2} \, d^{MNP} \vartheta_P \big) H_{\3 N} \; .
\end{equation}
The fermionic (gravitino and spin-$\frac12$) equations of motion are, in turn,
{\setlength\arraycolsep{0pt}
\begin{eqnarray}
  \label{eq:GravitinoEOM}
&&   \gamma^{\mu\nu\rho}\,D_\nu\psi_\rho{}^i   
+ \ft23 i\,  {\Omega}^{ij} \, {\cal P}_{\nu jk\ell m} \, \gamma^\nu\gamma^\mu \chi^{k \ell m} 
  -\tfrac{1}{4} H^{\rho\sigma M}  \Big( i {\cal V}_{M}{}^{ij} \gamma^{[\mu} \gamma_{\rho\sigma}\gamma^{\nu]} \psi_\nu{}^k\,  \Omega_{kj} 
- {\cal V}_{M \, jk} \gamma_{\rho\sigma} \gamma^{\mu}  \chi^{ijk} \Big) \nonumber\\[5pt]
&& \qquad   +3  i  \big( A_1{}^{ik} + 2 \, B^{ik} \big)  \Omega_{kj}\,  \gamma^{\mu\nu} \psi_\nu{}^j   
- \ft43  \,\Omega^{i\ell}  A_{2\, \ell, mnp}\, \gamma^\mu \chi^{mnp} 
- \tfrac{21}{2} \, B_{jk} \gamma^\mu \chi^{ijk} =0 \; , \\[20pt]
% 
% \,,
%
 \label{eq:FermionEOM}  &&   \slashed{D}  \chi^{ijk}  
 -\tfrac{i}{2} {\cal P}_\mu{}^{ijk \ell} \, \gamma^\nu \gamma^\mu \psi_\nu{}^m \,\Omega_{\ell m} 
+ \tfrac{3}{16} \, H^{\rho\sigma M }  {\cal V}_{M}{}^{[[ij}  \gamma^{\mu} \gamma_{\rho\sigma}\psi_\mu{}^{k]]} 
+ \tfrac{3i}{4} \, H^{\rho\sigma M }  {\cal V}_{M}{}^{ m [[i}  \gamma_{\rho\sigma} \chi^{jk]] n} \, \Omega_{mn} \nonumber \\[5pt]
&&   - \Omega_{mn}\, A_{2}{}^{m,ijk}\,   \gamma^{\mu}\psi_{\mu}{}^{n} 
-\tfrac98  \,  B^{[[ij} \gamma^\mu \, \psi_\mu{}^{k]]} 
+12i   A_{2}{}^{[[i,j|mp} \chi^{|k]] nq} \, \Omega_{mn} \Omega_{pq} \nonumber \\[5pt]
&& -3i A_1^{m[[i} \chi^{jk]] n } \Omega_{mn}  +18  i \,   B^{m[[i} \chi^{jk]] n } \Omega_{mn}   =0 \; .
\end{eqnarray}
}In these expressions, $M_{MN} = (\cV \cV^{\textrm{T}})_{MN} = \cV_M{}^{ik}\cV_N{}^{j\ell} \Omega_{ij} \Omega_{k\ell}$, with inverse $M^{MN}$, is the usual metric on $\mathrm{E}_{6(6)}/\mathrm{USp}(8)$, and the Hodge dual is taken w.r.t.~the spacetime metric $g_{\mu\nu}$. All quantities are appropriately covariantised. For example, $R_{\mu\nu}$ in the Einstein equation (\ref{eq:EinsteinEOM}) is a covariantised (due to the $A^M$-gauged trombone charge of $g_{\mu\nu}$) version of the usual Ricci tensor \cite{LeDiffon:2008sh}. The field strengths read
{\setlength\arraycolsep{0pt}
\begin{eqnarray}
\label{eq:H2Form} && H_\2^M \equiv dA^M + \tfrac12 X_{NP}{}^M A^N \wedge A^P +  \big(  Z^{MN} - \tfrac{15}{2} \, d^{MPN} \vartheta_P \big) B_N \; , \\[5pt]
\label{eq:H3Form} && H_{\3 M}  \equiv DB_{M} + d_{MNP} A^N \wedge dA^P + \tfrac13 \, d_{MNP} \,   X_{QR}{}^P A^N \wedge A^Q \wedge A^R \; .
\end{eqnarray}
}For both bosonic and fermionic fields, the covariant derivatives are built using the appropriate $\mathbb{R}^+ \times \mathrm{E}_{6(6)}$ representation $\bm{r}_w$ of the embedding tensor (\ref{eq:XSymbolsGen}), for example, 
\begin{equation} \label{eq:CovDerM}
DM_{MN} \equiv dM_{MN} -2 \,  A^P \,  \big( \Theta_{P}{}^\alpha +  \tfrac92 (t^\alpha)_{P}{}^{Q} \,  \vartheta_{Q} \big)  (t_\alpha)_{(M}{}^{R} \, M_{N)R} \; .
\end{equation}
For the fermions, the covariant derivatives acquire further terms in the scalar-dependent connection $Q_i{}^j$, in the adjoint, $\bm{36}$, of USp$(8)$, see \cite{deWit:2004nw} for full details and equation (\ref{eq:DerSusyParam}) in the appendix for a concrete example. The scalar current ${\cal P}^M{}_N \equiv M^{MP} D M_{PN}$ also appears in these equations, manifestly projected in the fermion equations as ${\cal P}^{ijk \ell} = {\cal P}^{[[ijk \ell]]}$, to the $\bm{42}$ of USp$(8)$. See appendix \ref{sec:USp8Relations} for the relation between E$_{6(6)}$ and USp$(8)$ indices. Also in the fermionic equations, the usual definitions $ \slashed{D} \equiv \gamma^\mu D_\mu$ and $\gamma_{\mu_1 \cdots \mu_p} = \gamma_{[\mu_1} \cdots  \gamma_{\mu_p]}$ have been employed, where $\gamma_\mu$, with $\{ \gamma_{\mu} , \gamma_{\nu} \} = 2\, g_{\mu\nu}$, denote the usual Clifford algebra generators. The `fermion shifts'
\begin{equation} \label{eq:FermionShifts}
A_1^{ij} \equiv \tfrac23 \, \Omega_{k\ell} \, {\cal Z}^{ik, j \ell} \; , \quad 
A_2^{i,jk\ell} \equiv 3 \, {\cal Z}^{i [j, k \ell]} + \tfrac32 \, A_1^{i[j} \, \Omega^{k\ell]} \; , \quad 
B^{ij} \equiv {\cal V}^{-1 ij M } \vartheta_M \; , 
\end{equation}
in the $\overline{\bm{36}}$, $\overline{\bm{315}}$ and  $\overline{\bm{27}}$ of USp$(8)$, have also been used. The shifts $A_1^{ij}$, $A_2^{i,jk\ell}$ were defined in \cite{deWit:2004nw} as the USp$(8)$-irreducible components of ${\cal Z}^{ij, k \ell} \equiv \tfrac{1}{\sqrt{5}} \, {\cal V}_M{}^{ij} \, {\cal V}_N{}^{k\ell} \, Z^{MN}$, and the trombone-dependent $B^{ij}$ is new.

The scalar and embedding tensor-dependent quantity $V$ that appears in the Einstein equation (\ref{eq:EinsteinEOM}) takes on the form
\begin{equation} \label{eq:Cosmo}
V  = \tfrac{27}{4}  \, M^{MN} \vartheta_M \vartheta_N  
+   \tfrac{1}{60} \, M^{MN} \, \big(   M^{PQ} M_{RS} \Theta_{MP}{}^R \Theta_{NQ}{}^S 
 +5 \, \Theta_{MP}{}^Q \Theta_{NQ}{}^P \big) \; ,
\end{equation}
or, in terms of the fermion shifts (\ref{eq:FermionShifts}), 
\begin{equation} \label{eq:CosmoFS}
V  =  \tfrac13 \, A_{2 i,jk\ell} \, A_2^{i,jk\ell} -3 \, A_{1 ij} A_1^{ij} +\tfrac{27}{4} \, B_{ij} B^{ij} \; .
\end{equation}
This reduces to the scalar potential of \cite{deWit:2004nw} when $\vartheta_M = 0$. The quantity  $\tilde{V}_{MN}$ that features in the scalar equation (\ref{eq:scalarEOM}) reads in turn:
{\setlength\arraycolsep{2pt}
\begin{eqnarray}
 \tilde{V}_{MN} & =& 
\tfrac12  \, \vartheta_P M^{PQ} \Theta_{Q(M}{}^R M_{N)R} +\tfrac{1}{60} 
   \Big( M^{PQ} M_{RS} \, \big(  \Theta_{MP}{}^R \Theta_{NQ}{}^S 
  +   \Theta_{PM}{}^R \Theta_{QN}{}^S \big)   \nonumber   \\
 && 
+ 5 \, \Theta_{MP}{}^Q \,  \Theta_{NQ}{}^P 
- M^{PQ} M^{RS} M_{T(M} M_{N)U} \Theta_{PR}{}^{T} \Theta_{QS}{}^{U} 
\Big)\, .
\label{VAB_CT} 
\end{eqnarray}
Only for $\vartheta_M = 0$ are $V$ and $\tilde{V}_{MN}$ related through $\tilde{V}_{MN} =  \partial V / \partial M^{MN}$, a relation that allows one to integrate those equations into a Lagrangian in the tromboneless case. Finally, the scalar equation (\ref{eq:scalarEOM}) appears projected throughout with
\begin{equation} \label{eq:ProjCoset}
P_{MN}{}^{PQ} \equiv M_{R(M} \, \mathbb{P}^R{}_{N)}{}^P{}_S \, M^{QS} \; ,
\end{equation}
a projector to the coset $\textrm{E}_{6(6)}/\textrm{USp}(8)$ \cite{Berman:2019izh} that can be written in terms of the E$_{6(6)}$ adjoint projector $\mathbb{P}^M{}_{N}{}^P{}_Q$ defined below (\ref{eq:defX}).

Let us conclude this section by listing the supersymmetry transformations of general $D=5$ $\cN=8$ gauged supergravity. The supersymmetry transformation of the bosonic fields read \cite{deWit:2004nw}
{\setlength\arraycolsep{0pt}\begin{eqnarray}
  \label{eq:susy-trans-bosons}
&&  \delta  e_\mu{}^a  = \tfrac12 \bar \epsilon_i  \gamma^a  \psi_\mu{}^i \,,\nonumber\\[4pt]
&&  \delta {\cal V}_{M}{}^{ij} = 4 i\, {\cal V}_{M}{}^{kl} \,\Omega_{p[[k} \bar\chi_{\ell mn]]} \epsilon^{p} \,\Omega^{mi}\,\Omega^{nj} \,, \nonumber\\[4pt]
&&  \delta A_\mu{}^{M}    = 2 \, \Big[i\,\Omega^{ik}\,\bar\epsilon_k
  \psi_\mu{}^j + \bar\epsilon_k \gamma_\mu \chi^{ijk}\Big] \,  
      {\cal V}_{ij}{}^{M}  \,, \nonumber\\[4pt]
&&  \delta B_{\mu\nu\,M} = \tfrac{4}{\sqrt5}\, {\cal V}_M{}^{ij}
       \Big[2\,\bar\psi_{[\mu | \,i}\gamma_{| \nu]} \epsilon^k\,\Omega_{jk} 
      -i\, \bar\chi_{ijk}  \gamma_{\mu\nu}\epsilon^k \Big]  +2\, d_{MNP}\,
  A_{[\mu}{}^N\,\delta A_{\nu]}{}^P \, , 
\end{eqnarray}
}
as in the ungauged theory, while the fermions transform as
{\setlength\arraycolsep{0pt}\begin{eqnarray}
  \label{eq:susy-trans-fermions}
&&  \delta \psi_\mu{}^i = D_\mu \epsilon^i
+ \tfrac{i}{12}\big( \gamma_{\mu\nu\rho}\,H^{\nu\rho\,M} 
  -4 \, \gamma^\nu\,H_{\mu\nu}{}^{M} \big)  \, {\cal V}_{M}{}^{ij}  \,  \Omega_{jk} \,\epsilon^k   - i \,\gamma_\mu\,
\big( A_1{}^{ij} + 2  B {}^{ij} \big)  \, \Omega_{jk} \,\epsilon^k  \,, \nonumber\\[6pt]
&&   \delta \chi^{ijk}  = {} \tfrac{i}{2} \, \gamma^\mu\,{\cal P}_\mu{}^{ijk\ell}
  \,\Omega_{\ell m} 
  \,\epsilon^m  
    - \tfrac{3}{16}\gamma^{\mu\nu}\, H_{\mu\nu}{}^M {\cal V}_{M}{}^{[[ij} \,\epsilon^{k]]} 
 +  A_2{}^{\ell,ijk}\, \Omega_{\ell m}\,\epsilon^m + \tfrac98 \, B^{[[ij} \epsilon^{k]]}   ,   \qquad
\end{eqnarray}
}again up to higher-order fermion terms. The non-trombone terms in (\ref{eq:susy-trans-fermions}) have been directly imported from \cite{deWit:2004nw}, and the new trombone contributions have been determined as outlined above. See appendix \ref{sec:susyEoms} for the details.

Although a maximal supergravity with gauged trombone symmetry cannot have a Lagrangian, it may well happen that subsectors thereof are described themselves by a perfectly defined Lagrangian. This will be the case whenever local charges under the scaling symmetry $\mathbb{R}^+$ are not activated for the retained fields. By (\ref{eq:XSymbolsGen}) and the form of the covariant derivatives specified below that equation, this will happen if the set of gauge fields $A^M_{\mathrm{trunc}}$ retained in the truncated subsector obey
\begin{equation} \label{eq:TrombVecContr}
\vartheta_M A^M_{\mathrm{trunc}} =0 \; .
\end{equation}
Interestingly, the embedding tensor component $\vartheta_M$ responsible for the gauging of the trombone symmetry in the parent $\cN=8$ theory will typically still leave its imprint on other couplings of the truncated subsector, like the first term of the scalar potential (\ref{eq:Cosmo}). Section \ref{eq:AdSvac} will discuss an example of this situation.

%%%%%%%%%%%

\subsection{Vacuum conditions and mass matrices} \label{sec:MassMat}

%%%%%%%%%%%

Vacuum configurations are obtained for all fields set to zero, except for the metric and the scalars. The former needs to take on the form corresponding to a maximally symmetric spacetime (Minkowski, de Sitter or anti-de Sitter), while the  latter must be constant and subject to the constraint
\begin{equation} \label{eq:Vacuum}
P_{MN}{}^{PQ} \, \tilde{V}_{PQ} = 0 \; ,
\end{equation}
which follows from the scalar equation of motion (\ref{eq:scalarEOM}). With these requirements, all the field equations (\ref{eq:EinsteinEOM})--(\ref{eq:FermionEOM}) are satisfied, with $V$ in (\ref{eq:Cosmo}) serving the role of the cosmological constant.

The mass matrices of the various fields can be obtained by linearising the equations of motion around the vacuum conditions (\ref{eq:Vacuum}). In the bosonic sector, while the gauge field and tensor mass matrices are straightforward to obtain from (\ref{eq:vectorEOM}) and (\ref{eq:tensorEOM}), deriving the scalar mass matrix from (\ref{eq:scalarEOM}) requires some more work. Eventually, these mass matrices can be brought to the form:
{\setlength\arraycolsep{0pt}
\begin{eqnarray}
\label{eq:VectorMassMat} && (\cM_{\textrm{vector}}^2)_M{}^N = 
\tfrac16 \, M^{NT} M^{QS} M_{ RP} \big( \Theta_{MQ}{}^R + \tfrac92 \,  \mathbb{P}^R{}_Q{}^U{}_M \, \vartheta_U \big) \big( \Theta_{TS}{}^P -\tfrac{27}{2} \,  \delta_T^P \, \vartheta_S \big) \nonumber \\[4pt]
&& \qquad \qquad \qquad \; + \tfrac16 \, M^{NT} \big( \Theta_{MP}{}^Q + \tfrac92 \,  \mathbb{P}^Q{}_P{}^U{}_M \, \vartheta_U \big) \big( \Theta_{TQ}{}^P -\tfrac{27}{2} \,  \delta_T^P \, \vartheta_Q \big) \; , \\[10pt]
\label{eq:TensorMassMat}  && (\cM_{\textrm{tensor}})_M{}^N = \sqrt{\tfrac25} \, M_{MP} \,  \big( Z^{PN} -\tfrac{15}{2}  \,  d^{PNQ} \, \vartheta_Q \big) \; , \\[10pt]
\label{eq:ScalarMassMat}  && (\cM_{\textrm{scalar}}^2)_{MN}{}^{PQ} = 12 \, M_{R_1 (M} \mathbb{P}^{R_1}{}_{N)}{}^{R_2}{}_{S_1} \tilde{V}_{S_2R_2} \, M^{S_1T_1} M^{S_2T_2} \, P_{T_1 T_2}{}^{PQ} \\[4pt]
&& \qquad \qquad \quad  + 12 \,  P_{MN}{}^{R_1 R_2} \big( U_{R_1 R_2 \, S_1 S_2} + W_{R_1 R_2 \, S_1 S_2} + W_{S_1 S_2 \, R_1 R_2 }  \big) \, M^{S_1T_1} M^{S_2T_2} \, P_{T_1 T_2}{}^{PQ} \; ,   \nonumber 
\end{eqnarray}
}\newpage

\noindent with $\mathbb{P}^{M}{}_{N}{}^{P}{}_{Q}$, $\tilde{V}_{MN}$ and $P_{MN}{}^{PQ}$ respectively defined above (\ref{eq:ProjId}), in (\ref{VAB_CT}) and in (\ref{eq:ProjCoset}). I have also used the shorthands
{\setlength\arraycolsep{2pt}
\begin{eqnarray}
U_{MN \, PQ } & \equiv & \tfrac{1}{2} \Big( 
M_{MR} \, \vartheta_{P} \,  \Theta_{QN}{}^{R} 
-M_{TP} M_{QM} M^{RS} \vartheta_{R} \Theta_{SN}{}^{T}
 \Big) \; ,  \\[10pt]
W_{MN \, PQ } & \equiv & \tfrac{1}{60} \Big( 
M_{R_1 R_2} \, \Theta_{MP}{}^{R_1} \,  \Theta_{NQ}{}^{R_2} 
 -M^{R_1 R_2} \,M_{P S_1} \,M_{Q S_2} \, \Theta_{MR_1}{}^{S_1} \,  \Theta_{NR_2}{}^{S_2} \nonumber \\[5pt]
&& \quad -M^{R_1 R_2} \,M_{P S_1} \,M_{Q S_2} \, \Theta_{R_1M}{}^{S_1} \,  \Theta_{R_2N}{}^{S_2}  \nonumber \\[5pt]
&& \quad + M^{R_1 R_2} M^{S_1S_2} M_{MP} M_{NT_1} M_{QT_2} \Theta_{R_1 S_1}{}^{T_1} \Theta_{R_2  S_2}{}^{T_2} \Big) \; , 
\end{eqnarray}
}such that $\partial \tilde{V}_{MN}/\partial M^{PQ} =  U_{(MN) (PQ) } + W_{(MN) (PQ)} + W_{(PQ)  (MN) } $. The superscript in $\cM_{\textrm{vector}}^2$ and $\cM_{\textrm{scalar}}^2$ indicates that the eigenvalues of these mass matrices on a given vacuum provide squared masses since the vector, (\ref{eq:vectorEOM}), and scalar, (\ref{eq:scalarEOM}), equations of motion are second order. In contrast, the tensor equation of motion (\ref{eq:tensorEOM}) is first order, and the eigenvalues of the mass matrix $\cM_{\textrm{tensor}}$ accordingly correspond to linear masses of either sign. The latter comment also applies to the fermion mass matrices for similar reasons. Linearising the fermionic equations of motion (\ref{eq:GravitinoEOM}), (\ref{eq:FermionEOM}) one obtains the following expressions for the gravitino and spin-$1/2$ fermion mass matrices,
{\setlength\arraycolsep{-.7pt}
\begin{eqnarray}
\label{eq:GravitinoMassMat} && (\cM_{\textrm{gravitino}})^{ij}   = \tfrac{3}{\sqrt{2}} \big( A_1{}^{ij} + 2B^{ij} \big)  \; ,  \\[5pt]
\label{eq:FermionMassMat} && (\cM_{\textrm{spin-$\frac12$}})^{ijk}{}_{\ell mn}   =  9\sqrt{2} \big(  4 \,  \delta^{[[i}_{[[\ell} \Omega_{m|p|} \Omega_{n]]q} \, A_{2}{}^{j,k]]pq} + \,  \delta^{[[i}_{[[\ell} \delta^{j}_{m} \Omega_{n]]p} \, A_{1}{}^{k]]p} +6  \delta^{[[i}_{[[\ell} \delta^{j}_{m} \Omega_{n]]p} \, B^{k]]p}  \big) ,   \qquad 
\end{eqnarray}
}with suitable normalisations.

Except for the gravitino mass matrix (\ref{eq:GravitinoMassMat}), all other mass matrices above contain unphysical eigenvalues in a given vacuum. Fortunately, these are straightforward to identify as follows. Firstly, (\ref{eq:VectorMassMat}) and (\ref{eq:TensorMassMat}) respectively contain $27-\mathrm{rank} \, X_{MN}{}^P$ and $\mathrm{rank} \, X_{MN}{}^P$ unphysical zero eigenvalues for the reasons reviewed at the end of section \ref{sec:5DFields}. Incidentally, any massless gauge field will also manifest itself as a zero, though now physical, eigenvalue of (\ref{eq:VectorMassMat}). Secondly, although (\ref{eq:ScalarMassMat}) is formally a square matrix of size 378, it is actually coset-projected with (\ref{eq:ProjCoset}) and thus it always contains 336 unphysical zero eigenvalues at any given vacuum. Similarly to the gauge fields, any massless scalar will manifest itself as an additional, though physical, zero eigenvalue. Finally, for symmetry-breaking and supersymmetry-breaking vacua, (\ref{eq:ScalarMassMat}) and (\ref{eq:FermionMassMat}) will respectively contain Goldstone and Goldstino modes eaten by the gauge fields and gravitini that become massive. The Goldstone modes arise as additional zero eigenvalues of (\ref{eq:ScalarMassMat}). At least for the AdS vacua discussed in section \ref{eq:AdSvac} but presumably more generally, a Goldstino mode arises as an eigenvalue $M_{\textrm{Goldstino}}$ of (\ref{eq:FermionMassMat}) related to the corresponding gravitino mass $M_{\textrm{gravitino}}$ as $M_{\textrm{Goldstino}} = \frac53 \, M_{\textrm{gravitino}}$. This relation does not hold, however,  for `massless' gravitini in a supersymmetric AdS vacuum. Those have $L M_{\textrm{`massless' gravitino}} = \pm \frac32 $ in units of the AdS radius $L$, and do not undergo superHiggsing.

%%%%%%%%%%%%%%
%%%%%%%%%%%%%%

\section{A new family of $D=5$ $\cN=8$ gauged supergravities} \label{sec:TCSO}

%%%%%%%%%%%%%%
%%%%%%%%%%%%%%

I will now move on to classify the theories with gauge group contained in the maximal subgroup $\mathbb{R}^+ \times \mathrm{SL}(2,\mathbb{R}) \times \mathrm{SL}(6,\mathbb{R})$ of $\mathbb{R}^+ \times \textrm{E}_{6(6)}$. This classification is exhausted by a new family of supergravities with gauge group that I will denote by TCSO$(p,q,r;\rho)$. The latter is characterised by four non-negative integers, $p$, $q$, $r$, $\rho$, subject to the constraints
\begin{equation} \label{eq:intconst}
p+q+r = 6 \; , \quad \rho \leq \textrm{min} \{ r, 2\} \; ,
\end{equation}
has the semidirect product structure
\begin{equation} \label{eq:ProdTCSO}
\textrm{TCSO} (p,q,r;\rho) = \textrm{T} (r;\rho) \, \ltimes \, \textrm{CSO} (p,q,r) = \big( \textrm{T} (r;\rho) \times \textrm{SO} (p,q) \big)   \ltimes \mathbb{R}^{(p+q)r}  \; ,
\end{equation}
in terms of the familiar $\textrm{CSO} (p,q,r) = \textrm{SO} (p,q)  \ltimes \mathbb{R}^{(p+q)r}$ and a group $\textrm{T} (r;\rho)$ defined below, and dimension
\begin{equation} \label{eq:dimTCSO}
\textrm{dim} \, \textrm{TCSO} (p,q,r;\rho) = \tfrac12 \, (p+q)(p+q-1) + (p+q) \, r + 2 \rho 
+ \rho ( r - \rho ) \; ,
\end{equation}
always less than $27$ under the restrictions (\ref{eq:intconst}).

%%%%%%%%%%%%%%
\subsection{Gaugings contained in $\mathbb{R}^+ \times \mathrm{SL}(2,\mathbb{R}) \times \mathrm{SL}(6,\mathbb{R})$} \label{sec:RplusSL6SL2Gaugings}
%%%%%%%%%%%%%%

To see how this family of gaugings arises, it is convenient to split the E$_{6(6)}$ generators and embedding tensor into $\mathbb{R}^+ \times \mathrm{SL}(2,\mathbb{R}) \times \mathrm{SL}(6,\mathbb{R})$ representations using (\ref{eq:27Split})--(\ref{eq:351Split}), and determine the possible couplings. These are summarised in table \ref{tab:Gaugings}. The left and right tables therein contain the splitting of the non-trombone and trombone contributions, and the $\mathbb{R}^+$ charges have been omitted throughout. Gaugings contained in $\mathbb{R}^+ \times \mathrm{SL}(2,\mathbb{R}) \times \mathrm{SL}(6,\mathbb{R})$ must only activate generators in the adjoint representations, $({\bf 1},{\bf 35})$ and $({\bf 3},{\bf 1})$, of $\mathrm{SL}(6,\mathbb{R})$ and $\mathrm{SL}(2,\mathbb{R})$. One extreme case is that of tromboneless gaugings with $\Theta_M{}^\alpha \neq 0$, $\vartheta_M =0$, captured by the left table in isolation. These were analysed in \cite{deWit:2004nw}, where it was found that this type of gaugings must only involve the $({\bf 1},{\bf 21})$ component of the embedding tensor and thus reduce to the CSO$(p,q,r)$ gaugings of  \cite{Gunaydin:1984qu,Gunaydin:1985cu,Pernici:1985ju,Andrianopoli:2000fi}. The other extreme case, $\Theta_M{}^\alpha = 0$, $\vartheta_M \neq 0$, can be analysed by looking at the right table by itself. One concludes that this type of gaugings cannot be contained in $\mathbb{R}^+ \times \mathrm{SL}(2,\mathbb{R}) \times \mathrm{SL}(6,\mathbb{R})$, as either component, $({\bf 1},{\bf 15})$ or $({\bf 2},{\bf \overline{6}})$, of the embedding tensor not only activates the $\mathrm{SL}(6,\mathbb{R}) \times \mathrm{SL}(2,\mathbb{R})$  generators, but also the $\textrm{E}_{6(6)}$ generators in the $({\bf 2},{\bf 20})$. 

For general embedding tensors, $\Theta_M{}^\alpha \neq 0$, $\vartheta_M \neq 0$,  in the $\overline{\bm{351}} + \bm{27}$ of E$_{6(6)}$ one must inspect both tables \ref{tab:Gaugings} combined. Embedding tensors with $\mathrm{SL}(2,\mathbb{R}) \times \mathrm{SL}(6,\mathbb{R})$ components $({\bf 1},{\bf 21})$ and $({\bf 2},{\bf \overline{6}})$ are now allowed provided the couplings to the $({\bf 2},{\bf 20})$ generators coming from the left and right tables offset each other and cancel altogether. This can be achieved as follows. Let $\theta_{AB} = \theta_{(AB)}$ and $\xi^{xA}$ respectively denote the $({\bf 1},{\bf 21})$ and $({\bf 2},{\bf \overline{6}})$ components descending from $\Theta_M{}^\alpha$, and $\vartheta^{xA}$ the $({\bf 2},{\bf \overline{6}})$ components coming from $\vartheta_M$, with all other components vanishing. Here, $A=1, \ldots , 6$ and $x=1,2$ are fundamental $\mathrm{SL}(6,\mathbb{R})$ and $\mathrm{SL}(2,\mathbb{R})$ indices, as in appendix \ref{sec:E6Conventions}. Under these circumstances, the embedding tensor (\ref{eq:defX}) reduces to 
\begin{eqnarray} \label{eq:XSymbolsNoConst}
&& X_{AB} = 2 \, \theta_{C[A} \, t_{B]}{}^C + \tfrac14 \big( 8 \xi^{xC} + 3 \vartheta^{xC} \big) \, t_{xABC}   \; , \nonumber \\[5pt]
&& X^{xA} = -\vartheta^{xA} \, \mathbf{1} + \tfrac34 \big( 8 \xi^{xB} - \vartheta^{xB} \big) \, t_B{}^A - \tfrac14 \big( 40 \xi^{yA} + 3 \vartheta^{yA} \big)  \, t_y{}^x \; ,
\end{eqnarray}
\newpage 

\noindent with representation indices omitted for readibility. Choosing $\vartheta^{xA} = -\tfrac83 \xi^{xA}$, the components of $ X_{AB} $ along the $({\bf 2},{\bf 20})$ generators $t_{xABC}$ indeed cancel and (\ref{eq:XSymbolsNoConst}) is left with combinations of $\mathbb{R}^+ \times \mathrm{SL}(6,\mathbb{R}) \times \mathrm{SL}(2,\mathbb{R})$ generators only, as desired.

\begin{table}[]

%\centering

\resizebox{\textwidth}{!}{

\begin{tabular}{c| c c c c} %\hline
~  &~&~\\[-4mm]
$\Theta_M{}^\alpha$ &$({\bf 1},{\bf 15})$
&&&$({\bf 2},\overline{\bf 6})$
\\ \hline
~&~&~\\[-3.5mm]
$({\bf 1},{\bf 35})$
&$({\bf 1},{\bf 21}) + ({\bf 1},{\bf 105}) $
&&&$({\bf 2},\overline{\bf 6})+({\bf 2},\overline{\bf 84}) $
\\
$({\bf 3},{\bf 1})$
& ({\bf 3},{\bf 15}) 
&&&$({\bf 2},\overline{\bf 6})$
\\
$({\bf 2},{\bf 20})$
&$({\bf 2},\overline{\bf 6})+({\bf 2},\overline{\bf 84}) $
&&&$ ({\bf 3},{\bf 15}) +  ({\bf 1},{\bf 105}) $
\\ \hline
\end{tabular}

\qquad

\begin{tabular}{c| c c c c} %\hline
~  &~&~\\[-4mm]
$ (t^\alpha)_M{}^N \vartheta_N $ &$({\bf 1},{\bf 15})$
&&&$({\bf 2},\overline{\bf 6})$
\\ \hline
~&~&~\\[-3.5mm]
$({\bf 1},{\bf 35})$
&$({\bf 1},{\bf 15}) $
&&&$({\bf 2},\overline{\bf 6}) $
\\
$({\bf 3},{\bf 1})$
& --- 
&&&$({\bf 2},\overline{\bf 6})$
\\
$({\bf 2},{\bf 20})$
&$({\bf 2},\overline{\bf 6}) $
&&&$ ({\bf 1},{\bf 15}) $
\\ \hline
\end{tabular}

}

\caption{\footnotesize{ Couplings between gauge fields and $\textrm{E}_{6(6)}$ generators induced by the $\overline{\bm{351}}$ (left, taken from \cite{deWit:2004nw}) and $\bm{27}$ (right) components of the embedding tensor, branched out into $\textrm{SL}(2,\mathbb{R}) \times \textrm{SL}(6,\mathbb{R})$ representations. 
}\normalsize}
\label{tab:Gaugings}
\end{table}

Happily, the restriction $\vartheta^{xA} = -\tfrac83 \xi^{xA}$ is also compatible with the quadratic constraints (\ref{eq:QCs}). To see this, it is helpful to also split the latter under $\mathrm{SL}(2,\mathbb{R}) \times \mathrm{SL}(6,\mathbb{R})$ using (\ref{eq:27Split}), (\ref{eq:351Split}), (\ref{eq:1728Split}). Some algebra allows one to reduce (\ref{eq:QCs}) to the following relations, 
\begin{eqnarray} \label{eq:SL6SL2QCs}
&(\bm{1}, \overline{\bm{15}}) & : \; \epsilon_{xy} \big( 3 \, \vartheta^{x[A | } + 8\,  \xi^{x[A| } ) \, \vartheta^{y | B]} =0  \; , 
\qquad 
\epsilon_{xy} \big( 3 \, \vartheta^{x[A | } + 8\,  \xi^{x[A| } ) \, \xi^{y | B]} =0  \; ,
 \nonumber \\[5pt]
& (\bm{1}, \overline{\bm{21}}) & : \;  \epsilon_{xy} \,  \xi^{x(A|}  \vartheta^{y|B)} =0  \; , 
 \nonumber \\[5pt]
& (\bm{2}, \bm{6}) & : \;  \theta_{AB} \, \vartheta^{xB} =0  \; , 
\qquad \qquad \qquad\qquad \quad \;\;
\theta_{AB} \, \xi^{xB} =0 \; , 
 \nonumber \\[5pt]
& (\bm{2}, \bm{120}) & :  \;  \theta_{AB} \, \big( 3 \, \vartheta^{xC} + 8\,  \xi^{xC} ) -\tfrac27 \, \delta_{(A}^C \, \theta_{B)D} \, \big( 3 \, \vartheta^{xD} + 8\,  \xi^{xD} ) =0  \; , \nonumber \\[5pt]
& (\bm{3},  \overline{\bm{15}}) & : \;  \xi^{x [ A | } \, \vartheta^{y | B]  }  -\tfrac12 \, \epsilon^{xy} \,\epsilon_{zt} \, \xi^{z [ A | } \, \vartheta^{t | B]  }  =0   \; ,
\end{eqnarray}
in the indicated $\mathrm{SL}(2,\mathbb{R}) \times \mathrm{SL}(6,\mathbb{R})$ representations. The choice
\begin{equation} \label{eq:QCSols}
\vartheta^{xA} = -\tfrac83 \xi^{xA} \; , \quad \theta_{AB} \, \xi^{xB} = 0 \; ,
\end{equation}
solves all quadratic constraints (\ref{eq:SL6SL2QCs}) and reduces the gauge group generators (\ref{eq:XSymbolsNoConst}) to 
\begin{equation} \label{eq:XSymbols}
X_{AB} = 2 \, \theta_{C[A} \, t_{B]}{}^C \; , \qquad
X^{xA} = \tfrac{8}{3} \, \xi^{xA} \, \, \mathbf{1} +8 \,  \xi^{xB} \, t_B{}^A - 8\,  \xi^{yA} \, t_y{}^x \; .
\end{equation}
For completeness, note that, under these conditions, the only active $\mathrm{SL}(2,\mathbb{R}) \times \mathrm{SL}(6,\mathbb{R})$ components of the $Z^{MN}$ embedding tensor (\ref{eq:ZembTen}) are
\begin{equation}
(\bm{1}, \bm{21}) \; : \; Z_{xA \ yB} = -\sqrt{\tfrac52 } \, \epsilon_{xy} \, \theta_{AB} \; , \qquad 
(\bm{2}, \overline{\bm{6}}) \; : \; Z^{AB}{}_{xC} = 4 \sqrt{10} \, \epsilon_{xy} \, \xi^{y [A} \delta^{B]}_C \; .
\end{equation}
From now on, $\vartheta^{xA}$ will be traded with $\xi^{xA}$ via the first relation in (\ref{eq:QCSols}).  

To summarise, the generators (\ref{eq:XSymbols}) with (\ref{eq:QCSols}) obey all linear and quadratic constraints and thus generate a Lie algebra whose associated Lie group, which will be denoted $\textrm{TCSO} (p,q,r;\rho)$ as advertised above, is a subgroup of $\mathbb{R}^+ \times \mathrm{SL}(2,\mathbb{R}) \times \mathrm{SL}(6,\mathbb{R})$. This Lie algebra has generic commutation relations
\begin{eqnarray} \label{eq:Commutators}
& [ X_{AB} , X_{CD} ] = 4 \, \theta_{[A [C} X_{D]  B]} \; , \qquad
[ X^{xA} , X^{yB} ] = 8 \big( \xi^{xB} \, X^{yA} - \xi^{yA} \, X^{xB} \big) \; , \nonumber \\[6pt]
& [ X^{xA} , X_{BC} ] = 16 \, \xi^{xD} \delta^A_{[B} \, X_{C]D} \; ,
\end{eqnarray}
obtained by bringing (\ref{eq:XSymbols}) to (\ref{eq:LieAlg}). It is apparent from (\ref{eq:Commutators}) that $X_{AB}$ and $X^{xA}$ separately close into two different subalgebras, with the latter acting semidirectly on the former. Up to outer automorphisms, the total Lie algebra is exclusively characterised by the eigenvalues, $p$ positive, $q$ negative and $r$ vanishing, of the quadratic form $\theta_{AB}$, and the rank, $\rho$, of the (in general rectangular) matrix $\xi^{xA}$. By construction, these four integers are restricted as in (\ref{eq:intconst}). Thus, it is convenient to split the SL$(6, \mathbb{R})$ index $A=1, \ldots , 6$ as $A=(i, a)$ with $i=1, \ldots , p+q$ and $a= 7-r , 8-r , \ldots , 6$, so that $i$ and $a$ respectively run over $p+q$ and $r$ values. Finally, one can set without loss of generality and in a compatible way with the second relation in (\ref{eq:QCSols}), 
\begin{equation} \label{eq:Charges1}
\theta_{AB} = \big( \theta_{ij}  \, , \, \theta_{ia} = 0  \, , \, \theta_{ab} = 0   \big) \; , \quad 
\xi^{xA} = \big( \xi^{xi} =0 \, , \, \xi^{xa}   \big) \; ,
\end{equation}
with only non-vanishing components
\begin{equation} \label{eq:Charges2}
\theta_{ij} = g_1 \, \textrm{diag} \, \big( 1, \stackrel{p}{\ldots} , 1 , -1 , \stackrel{q}{\ldots} , -1  \big) \; , \quad 
\xi^{xa} =
\left\{
\begin{array}{lll}
\xi^{16} = -\tfrac18  \, g_2 & ,                        & \textrm{if $\rho = 1$,}  \\
\xi^{16} = -\xi^{25} = -\tfrac18 \, g_2   & ,     & \textrm{if $\rho = 2$,} 
 \end{array} \right. \; 
\end{equation}
and, of course, $\xi^{xa}=0$ if $\rho = 0$. The chosen normalisation of $\xi^{16}$ simplifies some expressions below, and $g_1$, $g_2$ are non-zero couplings, set to $g_1 = g_2=1$ in the remainder of this section.

Under (\ref{eq:Charges1}), the only relevant generators (\ref{eq:XSymbols}) are 
\begin{equation} \label{eq:XSymbolsTCSOpqrrho}
X_{ij} = 2 \, \theta_{k[i} \, t_{j]}{}^k \; , \quad
X_{ia} =  \theta_{ki} \, t_{a}{}^k \; , \quad
X^{xa} = \tfrac{8}{3} \, \xi^{xa} \, \, \mathbf{1} +8 \,  \xi^{xb} \, t_b{}^a - 8\,  \xi^{ya} \, t_y{}^x \; , 
\end{equation}
since $X_{ab} = 0 $ and $X^{xi} = 8 \, \xi^{xa} (\theta^{-1})^{ij} X_{ja}$.  The first two sets of $\tfrac12 \, (p+q)(p+q-1) + (p+q) \, r$ generators in (\ref{eq:XSymbolsTCSOpqrrho}) span the Lie algebra of CSO$(p,q,r)$, while the $2 \rho + \rho ( r - \rho )$ generators $X^{xa}$ generate the Lie algebra of a Lie group that I will denote T$(r;\rho)$. Altogether, the number of generators in (\ref{eq:XSymbolsTCSOpqrrho}) is (\ref{eq:dimTCSO}). The generic commutators (\ref{eq:Commutators}) display a semidirect action of T$(r;\rho)$ on CSO$(p,q,r)$, as indicated by the first equality of (\ref{eq:ProdTCSO}). The specifics of these algebras and their action needs to be discussed for each value of $\rho$ in turn. 

Before turning to that, it is worth mentioning that for the $\textrm{TCSO} (p,q,r;\rho)$ supergravity, the redundancy in the vector-tensor sector can be removed as follows. The E$_{6(6)}$ indices for the gauge fields and the two-forms split under $\mathrm{SL}(2,\mathbb{R}) \times \mathrm{SL}(6,\mathbb{R})$ similarly to the embedding tensor, namely,
\begin{equation} \label{eq:Vec2FormSplit}
A^M = (A^{AB}, A_{xA}) \; , \qquad 
B_M = (B_{AB}, B^{xA}) \; , \qquad 
A=1 , \ldots , 6 , \quad x=1,2 ,
\end{equation} 
and then further down, $A=(i, a)$, $i=1, \ldots , p+q$, $a= 7-r , 8-r , \ldots , 6$, as in (\ref{eq:Charges1}). Selecting the $\textrm{dim} \, \textrm{TCSO} (p,q,r;\rho)$ vectors $A^{ij}$, $A^{ia}$, $A_{xa}$ and the $27 -\textrm{dim} \, \textrm{TCSO} (p,q,r;\rho)$ two-forms $B_{ab}$, $B^{xi}$ as non-vanishing, and setting the rest to zero, resolves the redundancy.

%%%%%%%%%%%%%

\subsection{$\mathrm{TCSO} (p,q,r;0)$} \label{sec:TCSOpqr0}

%%%%%%%%%%%%%

If $\rho=0$ with $0 \leq r \leq 6$, necessarily $\xi^{xa}=0$, and thus $X^{xa}=0$ by (\ref{eq:XSymbolsTCSOpqrrho}). The group $\mathrm{T}(r;0)$ is therefore trivial, and $\mathrm{TCSO} (p,q,r;0) = \mathrm{CSO} (p,q,r)$ is generated by $X_{ij}$, $X_{ia}$ in (\ref{eq:XSymbolsTCSOpqrrho}). From (\ref{eq:Commutators}), the non-vanishing Lie brackets are
\begin{equation} \label{eq:CommutatorsCSO}
[ X_{ij} , X_{k\ell} ] = 4 \, \theta_{[i [k} X_{\ell]  j]} \; , \qquad
[ X_{ij} , X_{ka} ] = -2 \, \theta_{k[i} X_{j] a } \; ,
\end{equation}
and illustrate the well-known fact that $\textrm{CSO} (p,q,r) = \textrm{SO} (p,q)  \ltimes \mathbb{R}^{(p+q)r}$, with $\textrm{SO} (p,q)$ rotations and $\mathbb{R}^{(p+q)r}$ abelian translations respectively generated by $X_{ij}$ and $X_{ia}$. The group $\mathrm{CSO} (p,q,r)$ has dimension (\ref{eq:dimTCSO}) with $\rho =0$, and is non-trivial only for $0 \leq r \leq 5$. Notable instances include the $r=0$ case, $\mathrm{CSO} (p,6-p,0) = \mathrm{SO} (p,6-p)$, $0\leq p \leq 3$, with only the $X_{ij}$, $i=1 , \ldots , 6$, generators present; the $r=1$, $q=0$ case, $\mathrm{CSO} (5,0,1) \equiv \mathrm{ISO} (5) \equiv \mathrm{SO} (5) \ltimes \mathbb{R}^{5} $, with $\mathrm{ISO} (5)$ the five-dimensional Euclidean group; and the $r=5$ case, $\mathrm{CSO} (1,0,5) = \mathbb{R}^{5}$, with only the abelian $X_{ia}$, $i=1$, $a=1 , \ldots, 5$, generators present. These gaugings involve the $\overline{\bm{351}}$ component of the embedding tensor only, not the $\bm{27}$ trombone components, and have been dealt with at length in \cite{deWit:2004nw,Gunaydin:1984qu,Gunaydin:1985cu,Pernici:1985ju,Andrianopoli:2000fi}.

%%%%%%%%%%%%

\subsection{$\mathrm{TCSO} (p,q,r;1)$} \label{sec:TCSOpqr1}

%%%%%%%%%%%%

For $1 \leq r \leq 6$ and $\rho \equiv \mathrm{rank}  \, \xi^{xa}  = 1$, the index $a$ is either fixed to $a=6$ for $r=1$ or splits as $a=(\alpha , 6)$ with $\alpha$ running over $r-1$ values as $\alpha = 7-r , 8-r, \ldots , 5$ for $2 \leq r \leq 6$. Then one can take, without loss of generality, $\xi^{16} \neq 0$, $\xi^{1\alpha} =0$, $\xi^{x\alpha} =0$, $x=1,2$, as in (\ref{eq:Charges2}). With the latter choice, the only non-vanishing generators are, from (\ref{eq:XSymbolsTCSOpqrrho}), 
\begin{equation} \label{eq:XSymbolsTCSOpqr1}
\mathrm{CSO}(p,q,r) \, : \; X_{ij}  \; , \; X_{i6} \; , \; X_{i\alpha} \; , \qquad 
\mathrm{T}(r; 1) \, : \; X^{16}  \; , \; X^{26} \; , \; X^{1\alpha} \; ,
\end{equation}
with $X_{i\alpha}$ and $X^{1\alpha}$ only present for $2 \leq r \leq 6$. The Lie algebra of $\mathrm{TCSO} (p,q,r;1)$ is thus spanned by the generators (\ref{eq:XSymbolsTCSOpqr1}), and has dimension (\ref{eq:dimTCSO}) with $\rho =1$. It is itself composed of the two subalgebras, $\mathrm{CSO}(p,q,r)$ and $\mathrm{T}(r;1)$, with the generators indicated in (\ref{eq:XSymbolsTCSOpqr1}) and the latter acting semidirectly on the former as in the first equality in (\ref{eq:ProdTCSO}).

The structure of these groups is best discussed by looking at the Lie brackets. From (\ref{eq:Commutators}), the non-vanishing commutators are 
\begin{eqnarray} \label{eq:CommutatorsCSOpqr1}
& [ X_{ij} , X_{k\ell} ] = 4 \, \theta_{[i [k} X_{\ell]  j]} \; , \qquad
[ X_{ij} , X_{k6} ] = -2 \, \theta_{k[i} X_{j] 6 } \; , \qquad
[ X_{ij} , X_{k\alpha} ] = -2 \, \theta_{k[i} X_{j] \alpha } \; , \nonumber \\[6pt]
& [ X^{16} , X^{26}] = - X^{26} \; , \qquad 
[ X^{16} , X^{1\alpha}] =  X^{1\alpha} \; ,  \\[6pt]
& \quad [ X^{16} , X_{i6}] =  X_{i6} \; , \qquad \quad 
[ X^{1\alpha} , X_{i\beta}] =  \delta^\alpha_\beta \,  X_{i 6} \; . \nonumber
\end{eqnarray}
The first line in (\ref{eq:CommutatorsCSOpqr1}) corresponds to the CSO$(p,q,r)$ commutation relations (\ref{eq:CommutatorsCSO}), with now $1 \leq r \leq 6$ and the index $a$ there split as above, $a=(\alpha , 6)$. The second line corresponds to the commutators of $\textrm{T} (r;1)$. These single out $\textrm{T} (1;1) \equiv B_2$, spanned by $X^{16}$, $X^{26}$, as the two-dimensional Borel subgroup $B_2$ of upper triangular matrices of an SL$(2, \mathbb{R})$ strictly embedded in all three factors of $\mathbb{R}^+ \times \mathrm{SL}(6,\mathbb{R}) \times \mathrm{SL}(2,\mathbb{R})$ per (\ref{eq:XSymbolsTCSOpqrrho}) with (\ref{eq:Charges2}). The commutators (\ref{eq:CommutatorsCSOpqr1}) also show that the Lie algebra of $\textrm{T} (r;1)$ with $2 \leq r \leq 6$ is an extension of $\textrm{T} (1;1)$ with $r-1$ abelian generators $X^{1\alpha}$, and the former contains the latter as a subalgebra. Altogether, these commutators characterise the $(r+1)$-dimensional Lie algebra of $\textrm{T} (r;1)$ as indecomposable, non-semisimple, solvable and non-nilpotent. In terms of Mubarakzyanov's classification of low-dimensional real Lie algebras, the Lie algebras of the first few cases $\textrm{T} (1;1)$, $\textrm{T} (2;1)$, $\textrm{T} (3;1)$ respectively are $\mathfrak{g}_{2.1}$, $\mathfrak{g}_{3.4}$ (also Bianchi VI$_0$, or three-dimensional Poincar\'e) and $\mathfrak{g}_{4.5}$. The Lie algebra of $\textrm{T} (2;1)$ can be also regarded as the In\"on\"u-Wigner contraction of the SL$(2, \mathbb{R})$ of which $\textrm{T} (1;1)$ is the Borel subgroup. Finally, the third line in (\ref{eq:CommutatorsCSOpqr1}) describes the semidirect action of $\textrm{T} (r;1)$ on CSO$(p,q,r)$. Only the $\mathbb{R}^{(p+q)r}$ translations are acted upon and the SO$(p,q)$ rotations commute with $\textrm{T} (r;1)$. This observation leads to the refinement of the semidirect action specified by the second equality of (\ref{eq:ProdTCSO}).

The list of noteworthy groups in the TCSO$(p,q,r;1)$ class includes the 17-dimensional $\mathrm{TCSO}(5,0,1;1) = B_2 \ltimes \mathrm{ISO}(5) = \big( B_2 \times \mathrm{SO}(5) \big) \ltimes \mathbb{R}^5 $, which has already appeared, though not with that name, in \cite{Bhattacharya:2024tjw}. Here, $\mathrm{ISO} (5) \equiv \mathrm{SO} (5) \ltimes \mathbb{R}^{5} $ is generated by $X_{ij}$, $X_{i6}$, $i=1, \ldots, 5$ and $B_2$ by $X^{16}$, $X^{26}$. Let us also mention the 11-dimensional $\mathrm{TCSO}(1,0,5;1) = \mathrm{T}(5;1) \ltimes \mathbb{R}^5$. This group features no CSO rotations and has $\mathrm{T}(5;1)$ generated by $X^{16}$, $X^{26}$, $X^{1\alpha}$, $\alpha=2, \ldots , 5$, and $\mathbb{R}^5$ by $X_{i\alpha}$, $X_{i6}$, with $i=1$. Finally, let us mention the 7-dimensional $\mathrm{TCSO}(0,0,6;1) = \mathrm{T}(6;1)$, generated by $X^{16}$, $X^{26}$, $X^{1\alpha}$, $\alpha=1, \ldots , 5$, where the entire CSO$(0,0,6)$ subgroup trivialises.

%%%%%%%%%%%%

\subsection{$\mathrm{TCSO} (p,q,r;2)$} \label{sec:TCSOpqr2}

%%%%%%%%%%%%

When $\rho \equiv \mathrm{rank}  \, \xi^{xa}  = 2$ for $2 \leq r \leq 6$, the index $a$ is either $a \equiv x^\prime = 5,6$ for $r=2$, or splits as $a=(\alpha , x^\prime)$ with $x^\prime =5,6$ and $\alpha$ running over $r-2$ values as $\alpha = 7-r , 8-r, \ldots , 4$ for $3 \leq r \leq 6$. Although the indices $x=1,2$ and $x^\prime = 5,6$ formally take different values by an unfortunate notational accident, they both label the fundamental representation of $\textrm{SL}(2,\mathbb{R})$. More precisely, both $x$ and $x^\prime$ are in the fundamental of the $\textrm{SL}(2,\mathbb{R})$ diagonal subgroup of the explicit $\textrm{SL}(2,\mathbb{R})$ factor in $\mathbb{R}^+ \times \mathrm{SL}(6,\mathbb{R}) \times \mathrm{SL}(2,\mathbb{R})$ (labelled by $x$) and $\textrm{SL}(2,\mathbb{R}) \subset \textrm{SL}(r,\mathbb{R}) \subset \textrm{SL}(6,\mathbb{R})$ (labelled by $x^\prime$). Thus, the primes may be dropped and $x$, $x^\prime$ identified. The tensor $\xi^{xa}$ can therefore be taken without loss of generality\footnote{Any other $\rho =2$ choice, for example the unnatural $\xi^{xy} = \delta^{xy}$, leads to the same Lie algebra up to automorphisms.}  as $\xi^{xy}   \propto \epsilon^{xy}$ for $r=2$, supplemented with $\xi^{x \alpha} = 0 $, $\alpha = 7-r , 8-r, \ldots , 4$, when $3 \leq r \leq 6$, as in (\ref{eq:Charges2}). This choice leads, from (\ref{eq:XSymbolsTCSOpqrrho}), to non-vanishing generators 
\begin{equation} \label{eq:XSymbolsTCSOpqr2}
\mathrm{CSO}(p,q,r) \, : \; X_{ij}  \; , \; X_{ix} \; , \; X_{i\alpha} \; , \qquad 
\mathrm{T}(r; 2) \, : \; X^{xy}  \; , \; X^{x\alpha} \; ,
\end{equation}
with $X_{i\alpha}$ and $X^{x\alpha}$ only present for $3 \leq r \leq 6$. Thus, the Lie algebra of $\mathrm{TCSO} (p,q,r;2)$ has dimension (\ref{eq:dimTCSO}) with $\rho =2$ and is spanned by the generators (\ref{eq:XSymbolsTCSOpqr2}). These close into the subalgebras indicated therein, with the latter again acting on the former as in the first relation in (\ref{eq:ProdTCSO}).

The non-vanishing commutators are, from (\ref{eq:Commutators}),
\begin{eqnarray} \label{eq:CommutatorsCSOpqr2}
& [ X_{ij} , X_{k\ell} ] = 4 \, \theta_{[i [k} X_{\ell]  j]} \; , \qquad
[ X_{ij} , X_{kx} ] = -2 \, \theta_{k[i} X_{j] x }  \; , \qquad
[ X_{ij} , X_{k\alpha} ] = -2 \, \theta_{k[i} X_{j] \alpha } \; ,  \nonumber \\[6pt]
& [ X^{xy} , X^{zt} ] = - \big( \epsilon^{xt} \, X^{zy} - \epsilon^{zy} \, X^{xt} \big)  \; , \qquad
[ X^{xy} , X^{z\alpha}] =  \epsilon^{zy} \, X^{x\alpha} \; ,  \\[6pt]
& [ X^{xy} , X_{iz}] =  \epsilon^{xt} \, \delta^y_z \,  X_{it} \; , \qquad
[ X^{x\alpha} , X_{i\beta}] =  \epsilon^{xy} \, \delta^\alpha_\beta \,  X_{i y} \; . \nonumber
\end{eqnarray}
Here, the first line contains the CSO$(p,q,r)$ commutation relations (\ref{eq:CommutatorsCSO}), with $2 \leq r \leq 6$ and the index $a$ further split as $a=(\alpha , x)$ as discussed above. The second line includes the commutation relations of $\textrm{T} (r;2)$, which make it apparent that $\textrm{T} (2;2) = \textrm{GL} (2, \mathbb{R}) $ and $\textrm{T} (r;2) = \textrm{GL} (2, \mathbb{R}) \ltimes \mathbb{R}^{2 (r-2)}$ for $3 \leq r \leq 6$. The subgroup $\textrm{GL} (2, \mathbb{R}) \equiv \textrm{SO}(1,1) \times \textrm{SL} (2, \mathbb{R})$ of $\textrm{T} (r;2)$ is generated by 
\begin{equation} \label{eq:XSymbolsGL2}
\mathrm{SO}(1,1) \, : \; \tilde{X} \equiv  \epsilon_{xy} \, X^{xy} \; ,  \qquad 
\textrm{SL} (2, \mathbb{R}) \, : \; \tilde{X}^{xy} = X^{xy} - \tfrac12 \, \epsilon^{xy} \,   \epsilon_{zt} \, X^{zt} \; ,
\end{equation}
with $\textrm{SL} (2, \mathbb{R})$ embedded diagonally in the two $\textrm{SL} (2, \mathbb{R})$ factors discussed above. By (\ref{eq:CommutatorsCSOpqr2}), the abelian translations $X^{x\alpha}$ of $\textrm{T} (r;2)$, $3 \leq r \leq 6$, transform under the $\textrm{GL} (2, \mathbb{R})$ subgroup as $r-2$ copies of the $\bm{2}_{1}$ representation. Finally, the third line in (\ref{eq:CommutatorsCSOpqr2}) describes the semidirect action of $\textrm{T} (r;2)$ on CSO$(p,q,r)$. Again, $\textrm{T} (r;2)$ commutes with $\mathrm{SO}(p,q)$, as reflected by the second equality of (\ref{eq:ProdTCSO}). Only the $\mathbb{R}^{(p+q)r}$ translations $X_{iz}$, $X_{i\alpha}$ are acted upon: $X_{iz}$ by the GL$(2, \mathbb{R})$ subgroup in $(p+q)$ copies of the $\bm{2}_{1} $ representation and, for  $3 \leq r \leq 6$, $X_{i\alpha}$ by the $\mathbb{R}^{2(r-2)}$ translations $X^{x\alpha}$.

Some notable TCSO$(p,q,r;2)$ groups include the 18-dimensional $\mathrm{TCSO}(p,4-p,2;2) = \big( \textrm{GL} (2, \mathbb{R}) \times \textrm{SO} (p, 4-p) \big) \ltimes \mathbb{R}^8$, $0\leq p \leq 4$, with each factor respectively generated by $X^{xy}$, $X_{ij}$, $X_{ix}$, and the semidirect actions specified above. Let us also mention the 15-dimensional $\mathrm{TCSO}(1,0,5;2) = \big( \textrm{GL} (2, \mathbb{R}) \ltimes \mathbb{R}^6 \big) \ltimes \mathbb{R}^5$, which features no CSO rotations and is generated by $X^{xy}$, $X^{x\alpha}$, $X_{ix}$, $X_{i\alpha}$, $i=1$, $\alpha =2,3,4$. Finally, note $\mathrm{TCSO}(0,0,6;2) = \mathrm{T}(6;2) = \textrm{GL} (2, \mathbb{R}) \ltimes \mathbb{R}^8$, where the entire CSO subgroup becomes trivial.

%%%%%%%%%%%%%%
%%%%%%%%%%%%%%

\section{Supersymmetric AdS solutions of $\mathrm{TCSO}(5,0,1;1)$} \label{eq:AdSvac}

%%%%%%%%%%%%%%
%%%%%%%%%%%%%%

Let us now fix for concreteness $p=5$, $q=0$, $r=\rho=1$ in the notation of the previous section, and focus on the $\mathrm{TCSO}(5,0,1;1)$ gauging. I will start by showing that this theory as presented in section \ref{sec:TCSOpqr1} and the maximal supergravity constructed \cite{Bhattacharya:2024tjw} by dimensional reduction on the $D=11$ MN2 solution of \cite{Maldacena:2000mw}, are in fact the same though expressed  in different duality frames. As it turns out, the particular duality frame of \cite{Bhattacharya:2024tjw} is just a member of a larger family of frames informed by consistent $\cN=8$ reduction \cite{BKV2025} on the BBBW family of $D=11$ solutions of \cite{Bah:2011vv,Bah:2012dg}. A convenient subsector of the $\mathrm{TCSO}(5,0,1;1)$ theory will be shown, in section \ref{sec:SusyVac}, to indeed have a family of supersymmetric AdS solutions in correspondence with their eleven-dimensional BBBW counterparts.

%%%%%%%%%%%%%%

\subsection{A convenient duality frame and supergravity subsector} \label{sec:DualFrameAndSector}

%%%%%%%%%%%%%%

As reviewed at the end of section \ref{sec:5DFields}, two five-dimensional maximal supergravities are equivalent if there is an E$_{6(6)}$ transformation that relates their embedding tensors through (\ref{eq:TransXSymbol}). Such transformation indeed exists that brings the $\mathrm{TCSO}(5,0,1;1)$ embedding tensor as given in section \ref{sec:TCSOpqr1} above, to that of the maximal supergravity discussed in \cite{Bhattacharya:2024tjw}. More generally, consider the following family of E$_{6(6)}$ transformations,
\begin{equation} \label{eq:TransMat}
U_{p^\prime q^\prime} = e^{ \, g_2 \, g_1^{-1} \, \frac{1}{p^\prime+ q^\prime}  \big( p^\prime \,  t_{2126} -  q^\prime \, t_{2346} \big)}  \; ,
\end{equation}
parametrised by two arbitrary real numbers $p^\prime$, $q^\prime$. Here, $ g_1$ and  $g_2$ are fixed, non-vanishing coupling constants, and the notation for the E$_{6(6)}$ generators is again as in appendix \ref{sec:E6Conventions} with representation indices omitted. Combining (\ref{eq:TransMat}) with the $\mathrm{TCSO}(5,0,1;1)$ embedding tensor, (\ref{eq:XSymbols}), (\ref{eq:Charges1}), (\ref{eq:Charges2}) with $p=5$, $q=0$, $r=\rho=1$, via the r.h.s.~of (\ref{eq:TransXSymbol}), a new embedding tensor results that reads
{\setlength\arraycolsep{2pt}
\begin{eqnarray} \label{eq:XSymbolsTilde}
\tilde{X}_{AB} & =&  2 \, \theta_{C[A} \, t_{B]}{}^C + \theta_{AB}{}^C{}_D \, t_C{}^D -2 \,  \xi^{xCD}{}_{[A|} \, t_{x|B] CD}   \; ,  \\[5pt]
\tilde{X}^{xA}  & =&  \tfrac{8}{3} \, \xi^{xA} \, \, \mathbf{1} +8 \,  \xi^{xB} \, t_B{}^A - 8\,  \xi^{yA} \, t_y{}^x  +2 \,  \xi^{xAB}{}_C \, t_B{}^C   -\tfrac{1}{36} \,  \epsilon^{xy} \,  \epsilon^{BCDEFG} \, \theta_{EF}{}^A{}_G \, t_{y BCD} \; . \nonumber
\end{eqnarray}
}Here, $\theta_{AB}$ and $\xi^{xA}$ are the tensors in the $(\bm{1},\bm{21})$ and $(\bm{2} , \overline{\bm{6}})$ representations of $\textrm{SL}(2,\mathbb{R} ) \times \textrm{SL}(6,\mathbb{R} )$ given in (\ref{eq:Charges1}), (\ref{eq:Charges2}) with $p=5$, $q=0$, $r=\rho=1$, while $\theta_{AB}{}^C{}_D = \theta_{[AB}{}^C{}_{D]}$, $\theta_{AB}{}^C{}_C =0$ and $\xi^{xAB}{}_C =\xi^{x[AB]}{}_C$, $\xi^{xAB}{}_B=0$ are tensors in the $(\bm{1},\bm{105})$ and $(\bm{2} , \overline{\bm{84}})$ whose only non-vanishing components up to antisymmetric permutations are
\begin{eqnarray}
\theta_{12}{}^6{}_5 = 6  \tfrac{q^\prime}{p^\prime +q^\prime} g_2^2 g_1^{-1}  , \quad
\theta_{34}{}^6{}_5 = -6 \tfrac{p^\prime}{p^\prime +q^\prime} g_2^2 g_1^{-1} , \quad
\xi^{212}{}_{6} = -\tfrac{p^\prime}{p^\prime +q^\prime} g_2  , \quad
\xi^{234}{}_{6} =  \tfrac{q^\prime}{p^\prime +q^\prime} g_2 .\quad 
\end{eqnarray}
For $p^\prime = 1$, $q^\prime = 0$, these reduce to the embedding tensor components (5) of \cite{Bhattacharya:2024tjw}, with $\kappa_{\textrm{there}}=-1$, $g_{3 \mathrm{there}} = g_2$, thus establishing the advertised equivalence. Incidentally, the $\kappa_{\textrm{there}}=0$ gauging straightforwardly coincides with $\mathrm{TCSO}(5,0,1;0)=\mathrm{CSO}(5,0,1)$.

The $D=11$ MN2 solution of \cite{Maldacena:2000mw} is a distinguished member of the more general BBBW class of solutions \cite{Bah:2011vv,Bah:2012dg}. So is the MN1 solution of \cite{Maldacena:2000mw}. It turns out that the MN2 truncation result of \cite{Bhattacharya:2024tjw} can be further extended to the BBBW class. More concretely, $D=11$ supergravity can be truncated consistently on the $(p^\prime, q^\prime)$-dependent internal manifold of the BBBW family \cite{BKV2025}, at least for hyperbolic Riemann surface therein\footnote{The $D=11$ BBBW solutions need to have $p^\prime$, $q^\prime$ integer-quantised in the full M-theory \cite{Bah:2011vv,Bah:2012dg}. The present $D=5$ supergravity description affords to have those real, as declared below (\ref{eq:TransMat}).}. This reduction yields a $D=5$ $\cN=8$ gauged supergravity with embedding tensor (\ref{eq:XSymbolsTilde}). By the same argument as above, this supergravity is again $\mathrm{TCSO}(5,0,1;1)$, expressed in a duality frame that relates to the canonical one of section \ref{sec:TCSOpqr1} via the $(p^\prime, q^\prime)$-dependent transformation (\ref{eq:TransXSymbol}) with (\ref{eq:TransMat}). The $\mathrm{TCSO}(5,0,1;1)$-gauged supergravity is thus expected to contain a family of $\cN=2$ AdS vacua that uplifts to the BBBW solutions in eleven dimensions. The time-honoured strategy I will use to pin down this family of vacua is to search for it within a smaller sector of the $D=5$ $\cN=8$ $\mathrm{TCSO}(5,0,1;1)$ theory. 

A subsector that does the job is provided by the $D=5$ $\cN=2$ model discussed in section 5 of \cite{Cassani:2020cod}. This model was obtained in that reference by straight reduction from $D=11$, without passing through the intermediate $D=5$ $\cN=8$ $\mathrm{TCSO}(5,0,1;1)$-gauged supergravity. The model of \cite{Cassani:2020cod} must, of course, be a subsector of the latter. That this is indeed the case can be seen by replicating their construction from a purely  $D=5$ perspective. In this context, the $D=5$ $\cN=2$ sector in question arises by imposing a U$(1)_S$-structure on the $D=5$ $\cN=8$ theory  generated by \cite{Cassani:2020cod}
\begin{equation} \label{eq:U1Struc}
t_{p^\prime q^\prime} = -\tfrac{p^\prime}{p^\prime+q^\prime}  ( t_1{}^2 - t_2{}^1 ) +\tfrac{q^\prime}{p^\prime+q^\prime}  ( t_3{}^4 - t_4{}^3 ) - (\tilde{t}_1{}^2-\tilde{t}_2{}^1) \; .
\end{equation}
Here and in (\ref{eq:ScalarGen}), the generators are again as in appendix \ref{sec:E6Conventions}, only with tildes over the specific SL$(2,\mathbb{R})$ generators involved to distinguish them from their SL$(6,\mathbb{R})$ counterparts. Interestingly, this $\mathrm{U}(1)_S \subset \mathrm{USp}(8)$ is not contained in the $\mathrm{TCSO}(5,0,1;1) = B_2 \ltimes \mathrm{ISO}(5)$ gauge group of the $D=5$ $\cN=8$ supergravity, yet truncation to the $\mathrm{U}(1)_S$-invariant fields is guaranteed to yield a consistent subsector of it. See \cite{Guarino:2024gke,Pico:2025cmc} for a recent related discussion.

The subsector of interest retains all U$(1)_S$-invariant fields, and only U$(1)_S$-invariant fields, of the parent $D=5$ $\cN=8$ theory. Along with the metric, these are  the scalars in the coset $\mathrm{C}_{\mathrm{E}_{6(6)}}(G_S) / \mathrm{C}_{\mathrm{USp}(8)}(G_S)$, where $\mathrm{C}_{K}(G)$ denotes the commutant of $G \subset K$ inside $K$, as well as the vectors, two-forms, gravitini and fermions that obey the restrictions
\begin{eqnarray} \label{eq:InvarFields}
& A^M  (t_{p^\prime q^\prime})_M{}^N = 0 \; , \quad  
(t_{p^\prime q^\prime})_M{}^N B_N = 0 \; , \quad \nonumber  \\%[4pt]
& (t_{p^\prime q^\prime})_M{}^N (\Gamma^M)^i{}_j (\Gamma_N)^j{}_k \psi^k_\mu =0  \; , \quad  
(t_{p^\prime q^\prime})_M{}^N (\Gamma^M)^{[[i}{}_m (\Gamma_N)^j{}_n \chi^{k]]mn} =0\; ,
\end{eqnarray}
for all $p^\prime$, $q^\prime$. The symbols $(\Gamma^M)^i{}_j$ that appear in these equations are defined in appendix \ref{sec:USp8Relations}. Inside $\textrm{E}_{6(6)}$, the U$(1)_S$ generated by (\ref{eq:U1Struc}) commutes with a subgroup $\textrm{U}(1) \times \textrm{SO}(1,1)^2 \times \textrm{SU}(2,1) $ generated by 
\begin{eqnarray} \label{eq:ScalarGen}
& J = \tfrac{1}{2} \, \big( t_1{}^2 - t_2{}^1+ t_3{}^4 - t_4{}^3 \big) \; , \qquad 
B = - \tfrac{1}{\sqrt{2}} \, \big( t_1{}^1 + t_2{}^2 - t_3{}^3 - t_4{}^4 \big) \; , \nonumber \\[5pt]
& H_0 = t_1{}^1 + t_2{}^2 + t_3{}^3 + t_4{}^4 - 2\, t_5{}^5 - 2 \, t_6{}^6 \; , \nonumber \\[5pt]
& H_1 = t_5{}^5 - t_6{}^6 \; , \qquad 
H_2 = -t_1{}^2 + t_2{}^1 + t_3{}^4 - t_4{}^3 +\tilde{t}_1{}^2 -\tilde{t}_2{}^1 \; , \nonumber  \\[5pt]
& E_{11} = \tfrac{1}{\sqrt{2}} \big( t_{1146} + t_{2246} + t_{2136} - t_{1236} \big) \; , \qquad 
E_{12} =  \tfrac{1}{\sqrt{2}} \big(  t_{1136} - t_{2146} + t_{2236} + t_{1246} \big) \; ,  \nonumber \\[5pt]
& F_{11} = -\tfrac{1}{\sqrt{2}} \big(  t_{1135} - t_{2145} + t_{1245} + t_{2235} \big) \; , \qquad 
F_{12} = \tfrac{1}{\sqrt{2}} \big( t_{1145} + t_{2245} + t_{2135} - t_{1235} \big)  \; ,  \nonumber \\[5pt]
& E_2 = \sqrt{2} \, t_6{}^5 \; , \qquad F_2 = \sqrt{2} \,  t_5{}^6 \; .
\end{eqnarray}
Here, $J$ generates U$(1)$, $B$, $H_0$ generate the two factors of SO$(1,1)$, and the remaining generators correspond to SU$(2,1)$, with normalisations chosen so that they close into the SU$(2,1)$ Lie algebra as given in (A.8) of \cite{Larios:2019kbw}. Thus, there are six $\textrm{U}(1)_S$-invariant real scalars, which will be denoted $\varphi_0$, $\varphi_1$, $\phi$, $a$, $\zeta$, $\tilde{\zeta}$, taking values on $\textrm{SO}(1,1)^2 \times \textrm{SU}(2,1)/ \left( \textrm{SU}(2) \times \textrm{U}(1)  \right)$. A coset representative for this manifold in Iwasawa gauge is
\begin{equation} \label{eq:SugraSector}
\cV = e^{\varphi_0 H_0} \, e^{-\sqrt{2} \, \varphi_1 B}  \, e^{-\frac{1}{\sqrt{2}} ( a E_2 - \zeta E_{11} +\tilde{\zeta} E_{12} ) } \, e^{-\phi H_1} \; .
\end{equation}
In order to determine the remaining invariant fields, it is convenient to split the E$_{6(6)}$ indices for the bosons as in (\ref{eq:Vec2FormSplit}) and introduce an explicit basis for six-dimensional gamma matrices to deal with the fermions. Then, solving equations (\ref{eq:InvarFields}) reveals the presence in the retained subsector of three U$(1)_S$-invariant gauge fields, $A^{12}$, $A^{34}$, $A^{56}$, three U$(1)_S$-invariant two-forms, $B_{12}$, $B_{34}$, $B_{56}$, along with two gravitini and four spin-$1/2$ fermions. This field content is compatible with an $\cN=2$ supergravity model coupled to two vector-tensor multiplets and one hypermultiplet, as in \cite{Cassani:2020cod}.

Extending the U$(1)_S$-invariance analysis to the embedding tensor would reveal the components that survive the truncation and define the couplings in the reduced sector. Alternatively, all these couplings as well as the equations of motion for this sector can be found by bringing the embedding tensor (\ref{eq:XSymbolsTilde}) and the above U$(1)_S$-invariant fields, along with the associated metric $M = \cV \cV^{\textrm{T}}$, gauge field strengths (\ref{eq:H2Form}), (\ref{eq:H3Form}), covariant derivatives (\ref{eq:CovDerM}), etc., into the $\cN=8$ field equations (\ref{eq:EinsteinEOM})--(\ref{eq:FermionEOM}). This is the route I will follow, simply noting that the $\cN=8$ trombone component $\tilde{\vartheta}_M \sim ( \tilde{\vartheta}_{[AB]} = 0 ,  \tilde{\vartheta} = -\tfrac18 \tilde{\xi}^{xA} \neq 0)$ that can be read off from the embedding tensor (\ref{eq:XSymbolsTilde}) contracts to zero with the U$(1)_S$-invariant vectors identified above, as in (\ref{eq:TrombVecContr}). By the discussion at the end of section \ref{sec:Eoms}, the resulting U$(1)_S$-invariant sector does not involve trombone gaugings and is described itself by a Lagrangian, in agreement with \cite{Cassani:2020cod}. In my conventions, this reads
\begin{eqnarray}	\label{eq:Lagrangian}
{\cal L} &=&  R \, \textrm{vol}_5 - 12 ( d\varphi_0 )^2   - 4 ( d\varphi_1 )^2  -  2(D\phi )^2 - \tfrac{1}{2} \, e^{4 \phi} \,   \big( Da +  \tfrac{1}{2}  ( \zeta D \tilde{\zeta} - \tilde{\zeta} D \zeta  ) \big)^2 \nonumber   \\[5pt]
&&  -  \tfrac{1}{2} \, e^{2 \phi} \,  ( D\zeta )^2    -  \tfrac{1}{2} \, e^{2 \phi} \,   (D\tilde{\zeta} )^2 - V \, \textrm{vol}_5  + \ldots   \ ,
\end{eqnarray}

\vspace{10pt}

\noindent where $ ( d\varphi )^ 2 \equiv d\varphi  \wedge * d\varphi $, etc., and the scalar potential is, setting for simplicity $g_1 = g_2 \equiv g$,
{\setlength\arraycolsep{2pt}
\begin{eqnarray} \label{eq:BBBWPot}
g^{-2} \, V &= &  \tfrac12 \, e^{4\phi + 8 \varphi_0} + 2e^{2\phi -4 \varphi_0}  - 2 e^{2( \phi + \varphi_0-\varphi_1) }- 2 e^{2( \phi + \varphi_0+ \varphi_1) } -4 e^{ -4 \varphi_0}  \nonumber \\[5pt]
&& +\tfrac12 \, e^{2\phi-4\varphi_0} ( \zeta^2 + \tilde{\zeta}^2 ) \big( e^{-4 \varphi_1} + e^{4 \varphi_1} +2 e^{2\phi} -2 \big) 
 \nonumber \\[5pt]
&& +\tfrac18 \, e^{4 (\phi- \varphi_0 + \varphi_1 ) } ( 1+ z - \zeta^2 - \tilde{\zeta}^2 )^2
+\tfrac18 \, e^{4 (\phi- \varphi_0 - \varphi_1 ) } ( 1 - z - \zeta^2 - \tilde{\zeta}^2 )^2 \; .
\end{eqnarray}
}It depends on the parameters $p^\prime$, $q^\prime$ only though the combination 
\begin{equation} \label{eq:defz}
z \equiv \frac{p^\prime-q^\prime}{p^\prime+q^\prime} \; ,
\end{equation}
and receives contributions from the parent $\cN=8$ trombone embedding tensor even if this sector by itself is tromboneless. The dots in (\ref{eq:Lagrangian}) include all other (gauge field, two-form and fermion) $\mathrm{U}(1)_S$-invariant fields in this subsector, and $D$ are covariant derivatives of the charged hypermultiplet scalars. None of these will be needed below.

Other $D=5$ $\cN=2$ or $\cN=4$ supergravity models \cite{MatthewCheung:2019ehr,Cassani:2019vcl,Faedo:2019cvr} that have been obtained by straight reduction from $D=11$ on the MN2, MN1 or BBBW solutions without passing through the intermediate $D=5$ $\cN=8$ TCSO$(5,0,1;1)$-gauged supergravity can be similarly recovered as subsectors of the latter. Indeed, the $D=5$ $\cN=4$ supergravity with three vector multiplets obtained in \cite{MatthewCheung:2019ehr,Cassani:2019vcl} by consistent truncation on MN2 is retrieved upon imposing a U$(1)_S$-structure (\ref{eq:U1Struc}) with $p^\prime =1$, $q^\prime =0$ on the $D=5$ $\cN=8$ TCSO$(5,0,1;1)$ supergravity, see \cite{Bhattacharya:2024tjw}. A U$(1)_S$-structure (\ref{eq:U1Struc}) with $p^\prime =q^\prime =1$ leads instead to the $D=5$ $\cN=2$ supergravity with four vector multiplets discussed in section 4 of \cite{Cassani:2020cod} (see also \cite{Faedo:2019cvr}), obtained in that reference by straight $D=11$  truncation on MN1.

%%%%%%%%%%%%%%%%%

\subsection{Supersymmetric vacua and mass spectra within gauged supergravity} \label{sec:SusyVac}

%%%%%%%%%%%%%%%%%

\begin{table}[]

%\centering

\resizebox{\textwidth}{!}{

\begin{tabular}{l | l | l } %\hline
\hline
\hline
\textbf{Field} & \textbf{MN1} & \textbf{MN2} \\[6pt]
\hline
Graviton & $0^{(1)}$ & $0^{(1)}$  \\[6pt]
\hline 
Gravitini & $\pm\tfrac52^{(1)}$, $\pm\tfrac32^{(1)}$, $\pm\tfrac{\sqrt{13}}{2}^{(2)}$ &
$\pm\tfrac52^{(2)}$, $\pm\tfrac32^{(2)}$  \\[6pt]
\hline 
Vectors & $8^{(2)}$, $6^{(1)}$, $3^{(2)}$,  $\left( \tfrac52 \pm \tfrac{\sqrt{13}}{2} \right)^{(4)}$, $0^{(4)}$ &
$8^{(5)}$, $3^{(8)}$, $0^{(4)}$   \\[6pt]
\hline 
Tensors & $\pm2^{(1)}$, $\pm \left( \tfrac12 + \tfrac{\sqrt{13}}{2} \right)^{(2)} $, $\pm \left( \tfrac12 - \tfrac{\sqrt{13}}{2} \right)^{(2)}$ &
$\pm3^{(1)}$, $\pm2^{(3)}$, $\pm1^{(1)}$  \\[6pt]
\hline
Spin-$\tfrac12$ & $\pm\tfrac{5}{2}^{(2)}$, $\pm\tfrac{3}{2}^{(4)}$, $\pm\tfrac{1}{2}^{(3)}$, $\pm \tfrac{\sqrt{13}}{2}^{(6)}$ $\pm \left( 1+ \tfrac{\sqrt{13}}{2} \right)^{(2)}$, $\pm \left( 1- \tfrac{\sqrt{13}}{2} \right)^{(2)}$, $\pm \left( \tfrac12 +\sqrt{7} \right)^{(1)}$, $\pm \left( \tfrac12 -\sqrt{7} \right)^{(1)}$ &
$\pm\tfrac72^{(2)}$, $\pm\tfrac52^{(10)}$, $\pm\tfrac32^{(8)}$, $  \pm\tfrac12^{(2)}$  \\[6pt]  
\hline
Scalars & $3^{(2)}$, $-3^{(6)}$, $-4^{(3)}$, $\left( -\tfrac12 \pm \tfrac{\sqrt{13}}{2} \right)^{(4)}$, $\left( 4 \pm 2\sqrt{7} \right)^{(1)}$, $0^{(8)}$ & 
$12^{(1)}$, $5^{(6)}$, $-3^{(6)}$, $-4^{(1)}$, $0^{(15)}$ \\[6pt]
\hline
\hline
\end{tabular}

\qquad 
}
\caption{\footnotesize{Squared masses (for gravitons, vectors and scalars) and linear masses (for gravitini, tensors, and spin-$1/2$ fermions) for the MN1 and MN2 solutions, in units of their respective AdS radii $L$, within $D=5$ $\cN=8$ $\textrm{TCSO}(5,0,1;1)$-gauged supergravity. A superscript in parenthesis indicates the multiplicity.
}\normalsize}
\label{tab:MN1MN2Masses}
\end{table}

For each value of $z$, the scalar potential (\ref{eq:BBBWPot}) reaches an AdS extremum at
{\setlength\arraycolsep{0pt}
\begin{eqnarray} \label{eq:BBBWvac}
& \textrm{BBBW: } \qquad  e^{12\varphi_0} = \frac{ z^2 \big( 2 + \sqrt{1+3z^2} \big)^2}{ \big( 3z-1 + \sqrt{1 + 3z^2 } \big) \big( 3z + 1 - \sqrt{1 + 3z^2}  \big) } \; , \qquad 
e^{4\varphi_1} = \frac{  3z-1 + \sqrt{1 + 3z^2 } }{ 3z + 1 - \sqrt{1 + 3z^2 } } \; ,  \\[5pt]
& \qquad \qquad e^{-2\phi} = \tfrac12 +\tfrac14  \sqrt{1 + 3z^2 } \; , \quad  \zeta = \tilde{\zeta} =0 \; , \quad
L^2 = \left( \frac{ 9 z^2 -1 + (1+3z^2)^{3/2}}{4z^2} \right)^{2/3} g^{-2} . \qquad \nonumber
\end{eqnarray}
}The AdS radius, $L^2 = -12/V$ with $V$ in (\ref{eq:BBBWPot}) evaluated on the scalar values in  (\ref{eq:BBBWvac}), has also been given in this equation. Regarded as vacua of the full TCSO$(5,0,1;1)$ theory, these generically break the $\cN=8$ supersymmetry spontaneously down to $\cN=2$, and preserve a residual $\textrm{U}(1)^2 \equiv \mathrm{U}(1)_F \times \mathrm{U}(1)_R$ subgroup of the gauge group, embedded into $\mathrm{USp}(8) \subset \mathrm{E}_{6(6)}$ through either route
\begin{equation} \label{eq:GroupTh}
\textrm{USp}(8) \supset
\textrm{USp}(4) \times \textrm{USp}(4) \supset
\textrm{USp}(4)_{\textrm{diag}} \sim
\textrm{SO}(5) \supset
\begin{array}{c}
\textrm{SO}(4) \supset \textrm{SU}(2)_F \times  \textrm{U}(1)_R  \\[4pt]
\textrm{SO}(3) \times  \textrm{SO}(2)
 \end{array} \supset
\textrm{U}(1)^2 \; .
\end{equation}
In $\textrm{U}(1)^2 \equiv \mathrm{U}(1)_F \times \mathrm{U}(1)_R$, the $\mathrm{U}(1)_R$ factor is the $\cN=2$ R-symmetry and $\mathrm{U}(1)_F$ is flavour. For specific values of the parameter $z$, the supersymmetry, the bosonic symmetry, or both, are enhaced. For $z=0$, corresponding to $p^\prime = q^\prime$ through (\ref{eq:defz}), 
\begin{equation} \label{eq:MN1vac}
\textrm{MN1 : } \quad z=0 \; , \quad e^{2 \phi} = \tfrac43 \; , \quad 
\varphi_0 = \varphi_1 = \zeta = \tilde{\zeta} =0 \; , \quad L^2 = \tfrac94 \, g^{-2} \; ,
\end{equation}
the supersymmetry is still $\cN=2$, but the flavour symmetry is enhanced to $\mathrm{SU}(2)_F$, making this solution enjoy as its bosonic  symmetry the $\mathrm{SU}(2)_F \times \mathrm{U}(1)_R$ that appears in the upper line of (\ref{eq:GroupTh}). For $z=1$, {\it i.e.}~$q^\prime=0$,  
\begin{equation} \label{eq:MN2vac}
\textrm{MN2 : } \quad z=1 \; , \quad e^{12 \varphi_0 } = e^{4 \varphi_1 }  = 2 \; , \quad 
\phi = \zeta = \tilde{\zeta} =0 \; , \quad L^2 = 2^{4/3}  \, g^{-2} \; ,
\end{equation}
the supersymmetry gets enhanced to $\cN=4$, with the R-symmetry accordingly augmented to the $\mathrm{SU}(2) \times \mathrm{U}(1) \sim \mathrm{SO}(3) \times \mathrm{SO}(2)$ that appears in the lower line of (\ref{eq:GroupTh}) and no residual flavour. The AdS vacua (\ref{eq:MN2vac}), (\ref{eq:BBBWvac}), (\ref{eq:MN1vac}) respectively uplift to the indicated $D=11$ solutions, MN2 \cite{Bhattacharya:2024tjw}, BBBW and MN1 \cite{BKV2025}, all of them with hyperbolic Riemann surface therein.

\begin{table}[]

%\centering

\resizebox{\textwidth}{!}{

\begin{tabular}{l | l  } %\hline
\hline
\hline
\textbf{Multiplet} & \textbf{BBBW}  \\[6pt]
\hline \\[-10pt]
Graviton & $A_1\bar{A}_1[3; \tfrac12 , \tfrac12 ; 0 ]_0$  \\[5pt]
\hline \\[-10pt]
Gravitino & $ L\bar{A}_2\left[\tfrac{7}{2};\tfrac{1}{2},0;1\right]_0 
 \oplus
A_2\bar{L}\left[\tfrac{7}{2};0,\tfrac{1}{2};-1\right]_0 
 \oplus
L\bar{L} \left[\Delta_+ ; \tfrac12 , 0  ;0\right]_{1}  
\oplus
L\bar{L} \left[\Delta_- ; \tfrac12 , 0 ;0\right]_{-1} 
\oplus
L\bar{L} \left[\Delta_+ ;  0, \tfrac12  ;0 \right]_{1}  
\oplus
L\bar{L} \left[\Delta_- ; 0, \tfrac12 ;0 \right]_{-1} $
  \\[6pt]
\hline \\[-10pt] 
Vector &  $A_2\bar{A}_2 \left[2;0,0;0\right]_0 
\oplus
L\bar{L} \left[\Delta_0 ;0, 0 ;0\right]_{0} 
\oplus
L\bar{L} \left[\Delta_1 ;0, 0 ;1\right]_{2}  
\oplus
L\bar{L} \left[\Delta_1 ;0, 0 ;1\right]_{-2}  $    \\[6pt]
\hline \\[-10pt] 
Chiral & $L\bar{B}_1 \left[3 ;0, 0 ;2\right]_{0}  
\oplus
B_1\bar{L} \left[3 ;0, 0 ;-2 \right]_{0}  $
\\[6pt]
\hline
\hline\\[-10pt] 
{\it Definitions:} & 
$\Delta_{\pm} \equiv 1 + \frac{\pm 1 \mp \sqrt{1+3z^2} + \sqrt{2+14z^2 +2(z^2-1)\sqrt{1+3z^2}} }{2 \, |z| }$, $ \quad  \Delta_0 \equiv 1 + \frac{\sqrt{4+13z^2 -4\sqrt{1+3z^2}} }{ |z| }$, $\quad \Delta_1 \equiv 1 + \frac{\sqrt{2 \big( 1+z^2 +(z^2-1)\sqrt{1+3z^2}} \big) }{ |z| }$.
\\[6pt]
\hline
\end{tabular}

\qquad 
}
\caption{\footnotesize{Spectrum of $\textrm{SU}(2,2| 1) \times \mathrm{U}(1)_F$ supermultiplets for the hyperbolic BBBW solution at fixed $z$, $z \neq 0, 1$, within $D=5$ $\cN=8$ $\textrm{TCSO}(5,0,1;1)$-gauged supergravity. A subscript denotes the $\mathrm{U}(1)_F$ charge.
}\normalsize}
\label{tab:BBBWMultiplets}
\end{table}

The masses of the various $D=5$ $\cN=8$ fields on these vacua can be computed by bringing the scalar vacuum values (\ref{eq:BBBWvac}), (\ref{eq:MN1vac}) or (\ref{eq:MN2vac}) and the embedding tensor (\ref{eq:XSymbolsTilde}) to the mass matrices (\ref{eq:VectorMassMat})--(\ref{eq:ScalarMassMat}), (\ref{eq:GravitinoMassMat}), (\ref{eq:FermionMassMat}). The diagonalisation results, for MN1 and MN2 only for brevity, have been listed in table \ref{tab:MN1MN2Masses}. The table includes only physical modes, as all Goldstone and spurious modes have been removed as explained at the end of section \ref{sec:MassMat}. For completeness, the massless graviton has been added to the table by hand.

These spectra arrange themselves, as they must, in representations of the residual supersymmetry groups: $\textrm{SU}(2,2| 2)$ for MN2, $\textrm{SU}(2,2| 1) \times \textrm{SU}(2)_F$ for MN1 and $\textrm{SU}(2,2| 1) \times \textrm{U}(1)_F$ for generic BBBW. Incidentally, this provides a reassuring crosscheck both on the 
\newpage 
\noindent field equations and on the mass matrices derived in sections \ref{sec:Eoms} and \ref{sec:MassMat}. In order to see this, one must firstly translate the masses $ML$ recorded in table \ref{tab:MN1MN2Masses}, referred to the respective AdS radii $L$, to conformal dimensions $\Delta$ via the usual formulae,
\begin{eqnarray}
& \textrm{gravitons, scalars:} \quad  \Delta (\Delta-4) = M^2L^2 \; , \qquad  
\textrm{vectors:} \quad  (\Delta -1) (\Delta-3) = M^2L^2 \; , \nonumber \\[5pt]
& \textrm{two-forms, gravitini, spin$-\frac12$ fermions:} \quad  \Delta = 2+ |ML| \; .
\end{eqnarray}
Secondly, the charges of all mass eigenstates under the relevant bosonic symmetry groups must be computed by branching the representations quoted in table \ref{tab:5DSummary} under the route in (\ref{eq:GroupTh}) appropriate to each solution. Finally, supermultiplets must be filled out of states with common flavour charge and appropriate patterns of dimensions and R-charges. The result of this lenghty exercise for the vacuum (\ref{eq:MN2vac}) that uplifts to the MN2 solution was already reported in (9) of \cite{Bhattacharya:2024tjw}. For the solutions (\ref{eq:BBBWvac}) and (\ref{eq:MN1vac}) that respectively uplift to generic BBBW and MN1, the supermultiplet structure is summarised in tables \ref{tab:BBBWMultiplets} and \ref{tab:MN1Multiplets}, respectively. My notation for the $\mathfrak{su}(2,2|1)$ supermultiplets is as in \cite{Cordova:2016emh}, with states  here, $[\Delta; j_1 , j_2 ; r] $, and there, $[j ; \bar{j}]_\Delta^{(r)}$, related by $\Delta_ {\textrm{there}}= \Delta_{\textrm{here}} $, $j_ {\textrm{there}}= 2 j_{1\textrm{here}} $, $\bar{j}_ {\textrm{there}}= 2 j_{2\textrm{here}} $, $r_ {\textrm{there}}= r_{\textrm{here}} $. For quick reference, the tables also quote the top spin component of each multiplet: graviton, gravitino, vector or spin-$1/2$, with the latter indicated as `Chiral'.

\begin{table}[]

%\centering

\resizebox{\textwidth}{!}{

\begin{tabular}{l | l  } %\hline
\hline
\hline
\textbf{Multiplet} & \textbf{MN1}  \\[6pt]
\hline \\[-10pt]
Graviton & $A_1\bar{A}_1[3; \tfrac12 , \tfrac12 ; 0 ] \otimes [0]$  \\[5pt]
\hline \\[-10pt]
Gravitino & $ L\bar{A}_2\left[\tfrac{7}{2};\tfrac{1}{2},0;1\right] \otimes [0]   
\; \oplus \;
A_2\bar{L}\left[\tfrac{7}{2};0,\tfrac{1}{2};-1\right] \otimes [0]  
\; \oplus \;
L\bar{L} \left[  1 +\tfrac{\sqrt{13}}{2}   ; \tfrac12 , 0 ;0\right] \otimes [\tfrac12]   
\; \oplus \;
L\bar{L} \left[  1 +\tfrac{\sqrt{13}}{2}   ;0, \tfrac12 ;0\right] \otimes [\tfrac12]  $
  \\[6pt]
\hline \\[-10pt] 
Vector &  $A_2\bar{A}_2 \left[2;0,0;0\right] \otimes [1]
\; \oplus \;
L\bar{L}\left[1+\sqrt{7};0,0;0\right] \otimes [0] $    \\[6pt]
\hline \\[-10pt] 
Chiral & $ L\bar{B}_1 \left[3 ;0, 0 ;2\right] \otimes [1]
\; \oplus \;
B_1\bar{L} \left[3 ;0, 0 ;-2 \right] \otimes [1]  $
\\[6pt]
\hline
\hline
\end{tabular}

\qquad 
}
\caption{\footnotesize{Spectrum of $\textrm{SU}(2,2| 1) \times \mathrm{SU}(2)_F$ supermultiplets for the MN1 solution, within $D=5$ $\cN=8$ $\textrm{TCSO}(5,0,1;1)$-gauged supergravity. The notation $\otimes [s]$ denotes the $\mathrm{SU}(2)_F$ representation of spin $s$.
}\normalsize}
\label{tab:MN1Multiplets}
\end{table}

Tables \ref{tab:BBBWMultiplets} and \ref{tab:MN1Multiplets} and equation (9) of \cite{Bhattacharya:2024tjw} reveal interesting patterns for these spectra. Some multiplets have superconformal primary dimensions independent of $z$. Thus, they are present, with adjustments in their flavour representations, for all members of the BBBW family, including MN1 and MN2. These include the expected stress-tensor, $A_1\bar{A}_1$, and flavour-current, $A_2\bar{A}_2$, multiplets, along with gravitino multiplets, $L\bar{A}_2$, $A_2\bar{L}$, and chiral multiplets, $L\bar{B}_1$, $B_1 \bar{L}$. For the charges indicated in the tables, these gravitino and chiral multiplets contain massless scalars: $\bm{1}_{0} +\bm{1}_{0} +\bm{3}_{0} +\bm{3}_{0}$ for MN1, in representations of $\textrm{SU}(2)_F \times \mathrm{U}(1)_R$ and in agreement with table \ref{tab:MN1MN2Masses}, and four for generic BBBW, all of them neutral under $\textrm{U}(1)_F \times \mathrm{U}(1)_R$. All other multiplets are long for generic BBBW and MN1, and their dimensions depend on $z$. Partial results on the spectrum of these solutions within gauged supergravity have been previously reported in \cite{Szepietowski:2012tb,MatthewCheung:2019ehr,Faedo:2019cvr,Cassani:2020cod}. For MN1 in particular, the spectrum given in \cite{Cassani:2020cod} contains the supermultiplets $A_1\bar{A}_1[3; \tfrac12 , \tfrac12 ; 0 ] \otimes [0]   \, \oplus \,A_2\bar{A}_2 \left[2;0,0;0\right] \otimes [1]   \, \oplus \, L\bar{L}\left[1+\sqrt{7};0,0;0\right] \otimes [0] $. The spectrum results given here are complete within five-dimensional gauged supergravity, in the sense that they have been obtained using the largest such type of $D=5$ theory.

%%%%%%%%%%%%%%
%%%%%%%%%%%%%%

\section{Discussion} \label{sec:Discussion}

%%%%%%%%%%%%%%
%%%%%%%%%%%%%%

The most general $D=5$ $\cN=8$ supergravity with local scaling symmetry has been specified building on \cite{deWit:2004nw,LeDiffon:2008sh}. Supergravities in this class do not admit a Lagrangian and must be formulated at the level of the field equations and supersymmetry transformations. The general gaugings contained in the maximal subgroup $\mathbb{R}^+ \times \mathrm{SL}(2,\mathbb{R}) \times \mathrm{SL}(6,\mathbb{R})$ of the ungauged duality group, $\mathbb{R}^+ \times \textrm{E}_{6(6)}$, have been described in detail and an interesting family, TCSO$(p,q,r;\rho)$, of gaugings has been uncovered. The TCSO$(p,q,r;\rho)$ family contains the usual CSO$(p,q,r)$ gauge groups \cite{Gunaydin:1984qu,Gunaydin:1985cu,Pernici:1985ju,Andrianopoli:2000fi} of tromboneless supergravity, and extend them to the case where trombone components of the embedding tensor are also present.

As shown in \cite{Bhattacharya:2024tjw} and section \ref{sec:DualFrameAndSector} above, one representative, TCSO$(5,0,1;1)$, in this class arises by consistent truncation of $D=11$ supergravity on the internal six-dimensional manifold $M_6$ corresponding to the MN2 sixteen-supercharge $\textrm{AdS}_5 \times M_6$ solution of \cite{Maldacena:2000mw}. From the five-dimensional perspective, this solution manifests itself as a half-maximal, supersymmetry breaking AdS vacuum of the $\cN=8$ TCSO$(5,0,1;1)$ theory. The TCSO$(5,0,1;1)$-gauged supergravity also turns out to arise by $D=11$ reduction on the internal manifold of the MN1, eight-supercharge AdS$_5$ solution of  \cite{Maldacena:2000mw}. More generally, the TCSO$(5,0,1;1)$ theory can also be obtained by reduction on the internal manifold of the BBBW, eight-supercharge AdS$_5$ solution of \cite{Bah:2011vv,Bah:2012dg}, for hyperbolic Riemann surface in all cases. Reference 
\cite{BKV2025} will elaborate on this observation, which lies beyond the strict five-dimensional scope of this paper. An indication of the relation of TCSO$(5,0,1;1)$ with all those $D=11$ solutions has nevertheless been given in section \ref{eq:AdSvac} above, where corresponding AdS vacua of that $D=5$ $\cN=8$ model have been found. The Kaluza-Klein (KK) spectrum of the MN1 and BBBW solutions, which has been given here only at level zero, {\it i.e.}, within the $D=5$ $\cN=8$ gauged supergravity,  will be also presented in \cite{BKV2025}. See \cite{Bhattacharya:2024tjw} for the KK spectrum on MN2.

More generally, it would be interesting to map the vacuum structure of this and other supergravities in the TCSO$(p,q,r;\rho)$ class by extending the methods of \cite{Krishnan:2020sfg,Bobev:2020ttg,Dallagata:2021lsc} to include trombone gaugings. It would also be of interest to elucidate the higher-dimensional origin, if any,  of these gauged supergravities beyond the TCSO$(5,0,1;0)$  and TCSO$(5,0,1;1)$ examples. The tromboneless $D=5$ $\cN=8$ $\textrm{CSO}(6,0,0) = \textrm{SO}(6)$-gauged supergravity is well-known to arise by consistent truncation of type IIB supergravity on the five sphere, $S^5$ \cite{Lee:2014mla,Ciceri:2014wya,Baguet:2015sma}. This resonates with the reduction \cite{Bhattacharya:2024tjw} of $D=11$ supergravity on MN2, whose internal geometry features a topological $S^4$. More generally, the tromboneless $D=5$ $\cN=8$ CSO$(p,q,r)$ gaugings are known to arise from consistent truncation on hyperboloids \cite{Hohm:2014qga,Baron:2014bya}. Those constructions should provide hints to determine the higher-dimensional origin of other supergravities in the TCSO$(p,q,r;\rho)$ class. 

\vspace{-2pt}

%%%%%%%%%%%%%%%
%%%%%%%%%%%%%%%

\section*{Acknowledgements}

%%%%%%%%%%%%%%%
%%%%%%%%%%%%%%%

I would like to thank Ritabrata Bhattacharya, Abhay Katyal and Mart\'\i n Pico for collaboration in related projects. This work was supported by NSF grant PHY-2310223.

%%%%%%%%%%%%%%%
%%%%%%%%%%%%%%%

\appendix

\addtocontents{toc}{\setcounter{tocdepth}{1}}

%%%%%%%%%%%%%%%
%%%%%%%%%%%%%%%

%%%%%%%%%%%%%%%
%%%%%%%%%%%%%%%

\section{$\mathrm{E}_{6(6)}$ and $\mathrm{USp}(8)$ conventions and useful relations} \label{sec:E6andUSp8Conventions}

%%%%%%%%%%%%%%%
%%%%%%%%%%%%%%%

This appendix summarises my conventions for $\textrm{E}_{6(6)}$ in the $\mathrm{SL}(2, \mathbb{R}) \times \mathrm{SL}(6, \mathbb{R})$ and USp(8) bases. It also collects some useful relations to deal with USp(8)-covariant expressions.

%%%%%%%%%%%%%%%

\subsection{$\mathrm{E}_{6(6)}$ conventions in the $\mathrm{SL}(2, \mathbb{R}) \times \mathrm{SL}(6, \mathbb{R})$ basis} \label{sec:E6Conventions}

%%%%%%%%%%%%%%%

Let $M = 1 , \ldots, 27$ and $\alpha = 1, \ldots , 78$ be fundamental and adjoint indices of $\textrm{E}_{6(6)}$, and $A=1, \ldots , 6$, $x=1,2$, fundamental indices of $\mathrm{SL}(6,\mathbb{R})$ and $\mathrm{SL}(2,\mathbb{R})$. Recall the branchings of some relevant $\textrm{E}_{6(6)}$ representations under $\mathrm{SL}(2, \mathbb{R}) \times \mathrm{SL}(6, \mathbb{R})$:
\begin{eqnarray}
\label{eq:27Split}
{\bf 27} &\rightarrow& ({\bf 1}, {\bf 15}) +
({\bf 2}, \overline{\bf 6}) \,,
\\
\label{eq:78Split}
{\bf 78} &\rightarrow&
({\bf 1},{\bf 35}) + ({\bf 3},{\bf 1})
+ ({\bf 2},{\bf 20})\,,
\\
\label{eq:351Split}
{\bf 351} &\rightarrow&
({\bf 1},\overline{{\bf 21}}) + ({\bf 3}, \overline{{\bf 15}}) + ({\bf 2},\bf 84)
+ ({\bf 2},{\bf 6}) + ({\bf 1}, \overline{{\bf 105}})\,,
\\
\label{eq:1728Split}
{\bf 1728} &\rightarrow&
({\bf 2},\overline{{\bf 6}}) + ({\bf 4},\overline{{\bf 6}}) + ({\bf 1},{\bf 15})
+ ({\bf 3},{\bf 15}) + ({\bf 2},\overline{\bf 84}) + ({\bf 1},{\bf 105})
\nonumber \\
 &&  + ({\bf 3},{\bf 105}) + ({\bf 2},\overline{\bf 120})  + ({\bf 2},\overline{\bf 210}) + ({\bf 1},{\bf 384}) \, .
\end{eqnarray}
Per (\ref{eq:78Split}), the $\textrm{E}_{6(6)}$ generators $t_\alpha$ (with representation indices omitted) split as $t_\alpha = ( t_A{}^B , \,  t_x{}^y , \, t_{xABC})$, where $t_A{}^B$ and $t_x{}^y$, with $t_A{}^A = t_x{}^x = 0$, respectively are the $\mathrm{SL}(6,\mathbb{R})$ and $\mathrm{SL}(2,\mathbb{R})$ generators, while $t_{xABC} = t_{x[ABC]}$ span the $({\bf 2},{\bf 20})$. By (\ref{eq:27Split}), the generators in the fundamental representation, $(t_\alpha)_M{}^N$, split into blocks as
\begin{equation}
\label{eq:GenE6Generic}
(t_\alpha)_M{}^N = 
\left(
\begin{array}{ll}
(t_\alpha)_{AB}{}^{CD} & (t_\alpha)_{AB \, yC} \\[3pt]
(t_\alpha)^{xA \, CD} & (t_\alpha)^{xA}{}_{yC}
\end{array}
\right) \; . 
\end{equation}
Altogether, the E$_{6(6)}$ generators can be written as
\begin{eqnarray}
\label{eq:GenE6Specific}
&& (t_A{}^B)_M{}^N = 
\left(
\begin{array}{ll}
(t_A{}^B)_{CD}{}^{EF} & (t_A{}^B)_{CD \, yG} \\[3pt]
(t_A{}^B)^{xC \, EF} & (t_A{}^B)^{xC}{}_{yG}
\end{array}
\right) = 
\left(
\begin{array}{cc}
4 \,  (t_A{}^B)_{[C}{}^{[E} \delta_{D]}^{F]} & 0 \\[3pt]
0 & - \delta^x_y \, (t_A{}^B)^C{}_G
\end{array}
\right) \; , \nonumber \\[12pt]
&&  (t_x{}^y)_M{}^N = 
\left(
\begin{array}{ll}
(t_x{}^y)_{AB}{}^{CD} & (t_x{}^y)_{AB \, zC} \\[3pt]
(t_x{}^y)^{tA \, CD} & (t_x{}^y)^{tA}{}_{zC}
\end{array}
\right) =
\left(
\begin{array}{cc}
0 & 0 \\[3pt]
0 & -(t_x{}^y)^{t}{}_{z} \, \delta_C^A
\end{array}
\right) \; ,  \\[12pt]
&& (t_{xABC})_M{}^N = 
\left(
\begin{array}{ll}
(t_{xABC})_{DE}{}^{FG} & (t_{xABC})_{DE \, zF} \\[3pt]
(t_{xABC})^{yD \, FG} & (t_{xABC})^{yD}{}_{zF}
\end{array}
\right) =
\left(
\begin{array}{cc}
0 & \epsilon_{xz}  \, \epsilon_{ABCDEF} \\[3pt]
6 \, \delta_x^y \, \delta_{[A}^D \delta_B^F \delta_{C]}^G & 0
\end{array}
\right) \; , \nonumber
\end{eqnarray}
where $\epsilon_{ABCDEF}$ and $\epsilon_{xy}$ are the usual Levi-Civita invariants of $\mathrm{SL}(6,\mathbb{R})$ and $\mathrm{SL}(2,\mathbb{R})$, 
\begin{equation} \label{eq:SLFun}
(t_A{}^B)_C{}^D = \delta_A^D \delta_C^B -\tfrac16 \, \delta_A^B \delta_C^D \; , \qquad 
(t_x{}^y)_z{}^t = \delta_x^t \delta_z^y -\tfrac12 \,  \delta_x^y \delta_z^t \; , \qquad 
\end{equation}
are the $\mathrm{SL}(6,\mathbb{R})$ and $\mathrm{SL}(2,\mathbb{R})$ generators in the fundamental representation, and $(t_A{}^B)^D{}_C$ and $(t_x{}^y)^t{}_z$ their transpose matrices. The transpose, $(t^{xABC})^N{}_M$, of the $({\bf 2},{\bf 20})$ generators $(t_{xABC})_M{}^N$ is, in this basis,
\begin{equation} \label{eq:Trans220}
(t^{xABC})^N{}_M = \tfrac16 \, \epsilon^{xy}  \, \epsilon^{ABCDEF} \, (t_{yDEF})_M{}^N \; .
\end{equation}

With the above conventions, the non-vanishing E$_{6(6)}$ Lie brackets read
\begin{eqnarray} \label{eq:E6Comm}
& [ t_A{}^B , t_C{}^D ]  = \delta_A^D \, t_C{}^B -  \delta_C^B \, t_A{}^D \; , \qquad
[ t_x{}^y , t_z{}^t ]  = \delta_x^t \, t_z{}^y -  \delta_z^y \, t_x{}^t \; , \nonumber \\[10pt]
& [ t_A{}^B , t_{xCDE} ]  = -3\,  t_{x A[CD}  \, \delta^B_{E]}  + \tfrac12 \delta_A^B \, t_{xCDE} \; , \nonumber \\[10pt]
& [ t_x{}^y , t_{zABC} ]  = \epsilon_{xz} \epsilon^{yt} \, t_{tABC} - \tfrac12 \delta_x^y \, t_{zABC} \; , \nonumber \\[10pt]
& [ t_{xABC} , t_{yDEF} ]  = -\tfrac32 \epsilon_{xy} \big( \epsilon_{GABC[DE} \, t_{F]}{}^G + \epsilon_{GDEF[AB} \,  t_{C]}{}^G \big) - \epsilon_{ABCDEF} \, \epsilon_{z(x } t_{y)}{}^z \; .
\end{eqnarray}
Finally, the non-vanishing components of the cubic invariant $d_{MNP}$, normalised so that (2.19), (2.20) of \cite{deWit:2004nw} hold, are
\begin{equation} \label{eq:CubicE6}
d_{AB \, CD \,  EF} = \tfrac{1}{\sqrt{10}} \epsilon_{ABCDEF} \; , \qquad
d_{AB}{}^{xC \, yD} = -\tfrac{6}{\sqrt{10}} \, \epsilon^{xy} \, \delta_A^{[C} \delta_B^{D]} \; .
\end{equation}
The components of $d^{MNP}$ are formally (\ref{eq:CubicE6}), only with upper and lower indices inverted.

%%%%%%%%%%%%%%%

\subsection{The $\mathrm{USp}(8)$ basis} \label{sec:USp8Relations}

%%%%%%%%%%%%%%%

It is often needed to bridge between E$_{6(6)}$ indices in the $\mathrm{SL}(2,\mathbb{R}) \times \mathrm{SL}(6,\mathbb{R})$ and the USp(8) bases. This situation arises, for example, to bring the coset representative $\cV_M{}^N$ of section \ref{eq:AdSvac}, obtained through exponentiation of the generators (\ref{eq:GenE6Specific}), to the notation $\cV_M{}^{ij}$ of section \ref{sec:GenGauge} on general formalism. I will also give some details on specific  USp$(8)$ representations pertaining to symplectic antisymmetrisations. This will be helpful for the calculations summarised in appendix \ref{sec:susyEoms}.

Let $A= 1 , \ldots , 6$ denote a vector index of $\mathrm{SO}(6) \subset \mathrm{SL}(6)$, borrowing the notation from appendix \ref{sec:E6Conventions}, and let $i=1, \ldots , 8$ be a fundamental index of USp$(8)$ as in the main text, though reiterpreted here as a Dirac spinor index of $\mathrm{SO}(6)$. Recall that the Clifford algebra of SO(6) has two inequivalent representations, differing on the symmetry or antisymmetry of the charge conjugation matrix. Let $(\Gamma_A)^i{}_j$ be the generators of the SO(6) Clifford algebra, $\{ \Gamma_A ,  \Gamma_B \} = 2 \, \delta_{AB}$, in the representation with antisymmetric charge conjugation matrix. The latter can be identified with the USp$(8)$-invariant matrix, $\Omega_{ij}$. In this representation, the symmetric and antisymmetric (in Dirac spinor indices) gamma matrices are
{\setlength\arraycolsep{0pt}
\begin{eqnarray} \label{SymmmetryCliff6}
&& \textrm{symmetric:} \; (\Omega \Gamma_a^{\mathrm{S}})_{ij} \equiv  \{ (\Omega \Gamma_7)_{ij}  \, , \;  i ( \Omega \Gamma_{AB})_{ij} \, , \;  i( \Omega \Gamma_{ABC})_{ij} \, \} , \quad a = 1, \ldots , 36, \qquad  \quad \\[5pt]
 \label{AntiSymmmetryCliff6}
&&  \textrm{antisymmetric:}  \; \Omega_{ij}  \, , \,  (\Omega \Gamma_M^{\mathrm{AS}})_{ij} \equiv  \{  ( \Omega \Gamma_{A})_{ij} \, , \, i (\Omega \Gamma_{A7})_{ij}   \, , \,  i (\Omega \Gamma_{AB7})_{ij} \}  , \, M = 1, \ldots ,  27, \qquad  \quad
\end{eqnarray}
}with $\Gamma_{A_1 \ldots A_n} \equiv \Gamma_{[A_1} \cdots \Gamma_{A_n]}$ as usual, $\Gamma_7$ the chirality matrix, and $\Gamma_{A_1 \ldots A_n 7} \equiv \Gamma_{A_1 \ldots A_n }\Gamma_{7}$. In (\ref{SymmmetryCliff6}), (\ref{AntiSymmmetryCliff6}), S and AS are just labels to emphasise the (anti)symmetry properties of the generators, while  $M = 1, \ldots ,  27$ and $a = 1, \ldots , 36$ are indices in the $\bm{27}$ and the adjoint of USp(8), as I will justify momentarily.

Tensors naturally occurring in E$_{6(6)}$ or USp$(8)$ representations can be written in the $\mathrm{SL}(2,\mathbb{R}) \times \mathrm{SL}(6,\mathbb{R})$ or the USp(8) bases by reducing them to the common $\mathrm{SO}(6) \sim \mathrm{SU}(4)$ subgroup, generated by $(\Gamma_{AB})^i{}_j$ in (\ref{SymmmetryCliff6}), and using gamma matrices appropriately. For example, the fundamental and the adjoint representations of E$_{6(6)}$ break down under $\mathrm{E}_{6(6)} \supset  \mathrm{USp}(8) \supset \mathrm{SU}(4) \times \mathrm{U}(1)$, with $\mathrm{U}(1)$ generated by $(\Gamma_{7})^i{}_j$ in (\ref{SymmmetryCliff6}), as
\begin{eqnarray}
  \label{eq:27E6toUSp8}
  && \textbf{27} \;
  \xrightarrow{\scriptscriptstyle \mathrm{USp}(8)} \;
   \textbf{27} 
  \xrightarrow{\scriptscriptstyle \mathrm{SO}(6)} \;
   \textbf{15}_0 + \textbf{6}_{2} + \textbf{6}_{-2} \; , \\
\label{eq:78E6toUSp8}
  && \textbf{78} \;
  \xrightarrow{\scriptscriptstyle \mathrm{USp}(8)} \;
   \textbf{36} + \textbf{42}
 \xrightarrow{\scriptscriptstyle \mathrm{SO}(6)} \;
   \big( \textbf{15}_0 + \textbf{10}_2+ \overline{\textbf{10}}_{-2} + \textbf{1}_0 \big) +
      \big( \textbf{20}^\prime_0 + \textbf{10}_{-2}+ \overline{\textbf{10}}_2 + \textbf{1}_{4}+ \textbf{1}_{-4} \big) \; . \nonumber
\end{eqnarray}
Thus, the symmetric and antisymmetric matrices in (\ref{SymmmetryCliff6}), (\ref{AntiSymmmetryCliff6}) respectively span the adjoint, $\bm{ 36}$, and the $\bm{ 27}$ of USp$(8)$. The composite connection $Q_i{}^j$ that enters the covariant derivative of the fermions and the coset representative $\cV_M{}^{ij}$ can then be written as
\begin{equation} \label{eq:QandVinUSp8}
Q_i{}^j = Q^a \, (\Gamma_a^{\mathrm{S}})_i{}^j \; , \qquad 
\cV_M{}^{ij} = \tfrac{1}{2\sqrt{2}} \, \cV_M{}^{N} \, (\Gamma_N^{\mathrm{AS}} \, \Omega)^{ij} \; ,
\end{equation}
with a suitable normalisation. The factors of $i$ in (\ref{AntiSymmmetryCliff6}) have been chosen so that, for $\cV_M{}^N$ real, the pseudoreality condition $\cV_{M \, ij} \equiv \Omega_{ik} \Omega_{j\ell}  \cV_M{}^{k\ell} = (\cV_{M}{}^{ij} )^*$ of \cite{deWit:2004nw} holds for $\cV_M{}^{ij} $ defined through (\ref{eq:QandVinUSp8}). %This convention makes the generators $(\Gamma_a^{\mathrm{S}})^i{}_j$ and $(\Gamma_M^{\mathrm{AS}})^i{}_j$ Hermitian, rather than anti-Hermitian. 
 Similarly, the scalar current ${\cal P}_\mu^{ijk\ell}$ lies in the $\bm{42}$ of USp$(8)$ and, according to the bottom line of (\ref{eq:27E6toUSp8}), can be expanded into SO(6)-covariant components as 
\begin{eqnarray}
{\cal P}_\mu^{ijk\ell} &=& {\cal P}_\mu^{AB} (\Gamma^C{}_{(A|7} \Omega)^{[[ ij}  (\Gamma_{|B)C 7 }  \Omega)^{k\ell]]}
+  {\cal P}_\mu^{ABC} (\Gamma_{[A| 7 } \Omega )^{ [[ ij}  (\Gamma_{|BC]7 }  \Omega)^{k\ell]]}
 \nonumber \\[4pt]
&& +  {\cal P}_\mu^0 (\Gamma^{A}\Omega)^{[[ ij }  (\Gamma_{A} \Omega)^{k\ell]]}  + \tilde{{\cal P}}_\mu^0  (\Gamma^{A}\Omega)^{[[ ij}  (\Gamma_{A7}  \Omega)^{k\ell]]}  \; .
\end{eqnarray} 

Here and in the main text, a double bracket has been used to denote symplectic antisymmetrisation, $T^{[[ i_1 i_2 \ldots i_n]]} \, \Omega_{i_1 i_2} =0$, for $T^{ i_1 i_2 \ldots i_n} = T^{[ i_1 i_2 \ldots i_n]}$, where single brackets indicate standard antisymmetrisation with weight one. It is  helpful to have expressions handy for these  in terms of regular antisymmetric tensors, particularly for the type of calculations sampled in appendix \ref{sec:susyEoms}. For example,
{\setlength\arraycolsep{2pt}
\begin{eqnarray}
U_{[[ij } V_{k\ell]]} & \equiv & U_{[ij } V_{k\ell]}
-\tfrac14 U_{[ij } \Omega_{k\ell ]} V_{mn } \Omega^{mn} 
-\tfrac14 \Omega_{[ij } V_{k\ell ]} U_{mn } \Omega^{mn} 
+ U_{[i|m|} V_{j|n|} \Omega_{k\ell ]} \Omega^{mn} \nonumber  \\[5pt]
&& + \tfrac{1}{24} \Omega_{[ij}\Omega_{k\ell]} U_{mn} V_{pq} \Omega^{mn} \Omega^{pq} 
- \tfrac{1}{12} \Omega_{[ij}\Omega_{k\ell]} U_{mn} V_{pq} \Omega^{mp} \Omega^{nq} \, , \\[8pt]
R_{[[ijk\ell]]} & \equiv & R_{ijk\ell} 
-\tfrac32 \Omega_{[ij} R_{k\ell ] mn } \Omega^{mn} 
- \tfrac18 \Omega_{[ij}\Omega_{k\ell]} R_{mnpq} \Omega^{mn} \Omega^{pq} \, , \\[8pt]
\label{eq:STSample}
T^{[[ij} x^{k]]} & \equiv & T^{[ij} x^{k]}
-\tfrac13 \Omega^{[ij} T^{k]m} \Omega_{mn} x^n \, , \\[8pt]
x^{[[i} y^j  z^{k]]} & \equiv & x^{[i} y^j  z^{k]} 
-\tfrac16 \Omega^{[ij} x^{k]} \Omega_{mn} y^m z^n
+\tfrac16 \Omega^{[ij} y^{k]} \Omega_{mn} x^m z^n
-\tfrac16 \Omega^{[ij} z^{k]} \Omega_{mn} x^m y^n \, , \qquad 
\end{eqnarray}
}where all tensors have been assumed to be antisymmetric, $U_{ij} = U_{[ij]}$, $V_{ij} = V_{[ij]}$, $R_{ijk\ell} = R_{[ijk\ell]}$, $T^{ij} = T^{[ij]}$, and only $T^{ij}$ has been further assumed to be symplectic-traceless, $T^{ij} = T^{[[ij]]}$.

%%%%%%%%%%%%%%%
%%%%%%%%%%%%%%%

\section{Supersymmetry and equations of motion} \label{sec:susyEoms}

%%%%%%%%%%%%%%%
%%%%%%%%%%%%%%%

I followed the procedure of \cite{LeDiffon:2008sh,LeDiffon:2011wt}, outlined in section \ref{sec:Eoms} above, to turn on the trombone-dependent contributions to the $\vartheta_M=0$ supersymmetry transformations given in \cite{deWit:2004nw}, and to the equations of motion that derive from the $\vartheta_M=0$ Lagrangian of the latter reference. This appendix contains some details on the derivation.

%%%%%%%%%%%%%%%

\subsection{Supersymmetry} \label{sec:susy}

%%%%%%%%%%%%%%%

The $\vartheta_M$--dependent contributions to the supersymmetry variations of the fermions are fixed by the USp(8) index structure to
\begin{equation} \label{eq:susy-trans-fermionsAnsatz}
\delta \psi_\mu{}^i = \ldots   -2i  \beta_1 \,  \gamma_\mu\,
  B {}^{ij}  \, \Omega_{jk} \,\epsilon^k  \,, \qquad
 \delta \chi^{ijk}  = \ldots
 +  \beta_2  \, B^{[[ij} \epsilon^{k]]} \, ,   \quad
\end{equation}
leaving only two unknown coefficients $\beta_1$ and $\beta_2$. In these expressions, $B^{ij}$ is defined  in (\ref{eq:FermionShifts}), and the dots stand for all non-trombone terms as given in (5.5) of \cite{deWit:2004nw}. These coefficients can be fixed by requiring the supersymmetry algebra to close into the bosonic symmetries specified in (5.9) of \cite{deWit:2004nw}. In the case at hand, it is enough to close the algebra on just the vielbein and the scalar coset representative, where  gauge transformations
\begin{equation} \label{eq:GaugeTrans}
\delta_\Lambda e_\mu{}^\alpha = \Lambda^M \, X_M^{\bm{1}_1} \, e_\mu{}^\alpha \; , \qquad 
 \delta_\Lambda \cV_N{}^{ij } = \Lambda^M \,   (X_M^{\bm{27}_0})_N{}^P \,   \cV_P{}^{ij}  \; ,
\end{equation}
with parameter \cite{deWit:2004nw}
\begin{equation} \label{eq:ParamGauge}
\Lambda^M = -2i \cV^{-1 ij M} \Omega_{jk} \, \bar{\epsilon}_{2i} \, \epsilon_1^k \; ,
\end{equation}
must be generated by the consecutive action of two supersymmetry transformations:
\begin{equation} \label{eq:CommSusy}
[ \delta_1 , \delta_2 ] \, e_\mu{}^\alpha = \delta_\Lambda e_\mu{}^\alpha \; , \qquad 
\cV^{-1 [[ij| N } \, [ \delta_1 , \delta_2 ] \, \cV_N{}^{| k\ell]] } = \cV^{-1 [[ij| N } \, \delta_\Lambda \cV_N{}^{| k\ell]] }  \; . 
\end{equation}
In the second equation here, the action of the supersymmetry algebra has been projected to the coset $\mathrm{E}_{6(6)}/\mathrm{USp}(8)$ by contraction with $\cV^{-1 ij N }$ and symplectic antisymmetrisation, in order to prevent a local USp(8) transformation from appearing on the r.h.s.

The l.h.s.~of the equations (\ref{eq:CommSusy}) can be computed by using in the fermionic, (\ref{eq:susy-trans-fermionsAnsatz}), and bosonic, (\ref{eq:susy-trans-bosons}), supersymmetry transformations followed by standard gamma matrix algebra and spinor bilinear manipulations. Bringing in the embedding tensor (\ref{eq:VariousX}) in the appropriate representations, the result does match (\ref{eq:GaugeTrans}) with parameter  (\ref{eq:ParamGauge}), provided the coefficients are fixed to $\beta_1 = 1$, $\beta_2 = \frac98$ as reported in (\ref{eq:susy-trans-fermions}) of the main text. This exercise is straightforward, though slightly tedious, for the vielbein, but requires some work for the scalar coset. The latter calculation involves using the identity (\ref{eq:ProjId}) in the definition of $(X_M^{\bm{27}_0})_N{}^P$, and then heavy use of various scalar coset identities including (4.10) and (4.26) of \cite{deWit:2004nw}. These manipulations allow one to show the identity
\begin{equation}
d_{MNR} \, d^{QPR} \, \cV^{-1 [[ij| N } \, \cV_P{}^{| k\ell]] } \, \vartheta_Q = -\tfrac12 \, B^{[[ij } \,  \cV_M{}^{k\ell]] } \; ,
\end{equation}
which comes in helpful in the intermediate steps of the calculation.

%%%%%%%%%%%%%%%

\subsection{Equations of motion} \label{sec:EOMs}

%%%%%%%%%%%%%%%

Having determined the supersymmetry variations, the strategy to determine the full set of equations of motion starts by first writing an ansatz for the fermionic field equations including trombone terms with unknown coefficients, and then vary it under supersymmetry. Variation generates the bosonic field equations including trombone terms with coefficients that depend on the above unknown parameters. It also generates spurious terms that must be forced to vanish. The latter requirement fixes the parameters and thus all coefficients in the fermionic and bosonic equations of motion. Again, this exercise needs to be performed in the case at hand by keeping track of the $\vartheta_M$-dependent terms only, as all remaining trombone-independent contributions follow from \cite{deWit:2004nw}. I will illustrate this process by fixing the trombone parameters in the fermionic equations. Fixing the analogue terms in the bosonic field equations is achieved by similar manipulations.

The starting point is the following ansatz for the fermionic equations of motion,
{\setlength\arraycolsep{0pt}
\begin{eqnarray}
  \label{eq:GravitinoEOMAnsatz}
&&   \gamma^{\mu\nu\rho}\,D_\nu\psi_\rho{}^i   
+ \ft23 i\,  {\Omega}^{ij} \, {\cal P}_{\nu jk\ell m} \, \gamma^\nu\gamma^\mu \chi^{k \ell m} 
  -\tfrac{1}{4} H^{\rho\sigma M}  \Big( i {\cal V}_{M}{}^{ij} \gamma^{[\mu} \gamma_{\rho\sigma}\gamma^{\nu]} \psi_\nu{}^k\,  \Omega_{kj} 
- {\cal V}_{M \, jk} \gamma_{\rho\sigma} \gamma^{\mu}  \chi^{ijk} \Big) \nonumber\\[5pt]
&& \qquad   +\tfrac32  i   \big( 2 A_1{}^{ik} + \lambda \, B^{ik} \big)  \Omega_{kj}\,  \gamma^{\mu\nu} \psi_\nu{}^j   
- \ft43 \,\Omega^{i\ell}  A_{2\, \ell, mnp}\, \gamma^\mu \chi^{mnp} 
+ \tfrac12  \, \zeta \, B_{jk} \gamma^\mu \chi^{ijk} =0 \; , \\[20pt]
% 
% \,,
%
 \label{eq:FermionEOMAnsatz}  &&   \slashed{D}  \chi^{ijk}  
 -\tfrac{i}{2} {\cal P}_\mu{}^{ijk \ell} \, \gamma^\nu \gamma^\mu \psi_\nu{}^m \,\Omega_{\ell m} 
+ \tfrac{3}{16} \, H^{\rho\sigma M }  {\cal V}_{M}{}^{[[ij}  \gamma^{\mu} \gamma_{\rho\sigma}\psi_\mu{}^{k]]} 
+ \tfrac{3i}{4} \, H^{\rho\sigma M }  {\cal V}_{M}{}^{ m [[i}  \gamma_{\rho\sigma} \chi^{jk]] n} \, \Omega_{mn} \nonumber \\[5pt]
&&   - \Omega_{mn}\, A_{2}{}^{m,ijk}\,   \gamma^{\mu}\psi_{\mu}{}^{n} 
-\tfrac34  \, \mu \,  B^{[[ij} \gamma^\mu \, \psi_\mu{}^{k]]} 
+12i   A_{2}{}^{[[i,j|mp} \chi^{|k]] nq} \, \Omega_{mn} \Omega_{pq} \nonumber \\[5pt]
&& -3i A_1^{m[[i} \chi^{jk]] n } \Omega_{mn}   -\tfrac12  i \, \nu  B^{m[[i} \chi^{jk]] n } \Omega_{mn}   =0 \; ,
\end{eqnarray}
}where $\lambda$, $\mu$, $\zeta$, $\nu$ are  unknown coefficients in front of all possible trombone-dependent terms. The form of these contributions is fixed by USp(8) covariance. All the remaining terms in (\ref{eq:GravitinoEOMAnsatz}) and (\ref{eq:FermionEOMAnsatz}) follow from variation of the Lagrangian (5.15) of \cite{deWit:2004nw} under $\bar{\psi}_{\mu i }$ and $\bar{\chi}_{ijk }$, respectively.

In order to fix these coefficients, it is enough to look at a few specific contributions obtained upon supersymmetry variation. Contributions of the schematic form $B {\cal D} \epsilon$ arise from variation under (\ref{eq:susy-trans-fermions}) of the first and sixth terms in (\ref{eq:GravitinoEOMAnsatz}), and also from the first and sixth terms in (\ref{eq:FermionEOMAnsatz}). Using standard gamma matrix manipulations, these terms respectively reduce to the l.h.s.'s of
\begin{equation} \label{eq:BDeps}
3i (\lambda -4\beta_1 ) B^{ij} \Omega_{jk} \gamma^{\mu\nu} D_\nu \epsilon^k = 0 \;  , \qquad
 \big( \mu -\tfrac43 \beta_2 ) \gamma^\mu B^{[[ i } D_\mu \epsilon^{k ]] } = 0 \; . 
\end{equation} 
Such contributions must be absent from the bosonic equations of motion and thus their vanishing must be enforced as indicated. Here and below, the supersymmetry variations of the fermions have been actually used in the form (\ref{eq:susy-trans-fermionsAnsatz}), in order to crosscheck the coefficients $\beta_1$, $\beta_2$ therein. 

Next, the supersymmetry variation of the gravitino equation (\ref{eq:GravitinoEOMAnsatz}) must yield the Einstein equation contracted with $\gamma_\mu$ and the gauge field equation affected by no gamma matrices. Terms affected by $\gamma_{\mu\nu}$ and $\gamma_{\mu\nu\rho}$ that are generated upon variation are thus spurious and must be forced to vanish. Contributions of the schematic form $B H^M_{\mu\nu} \gamma^{\mu\nu\rho}$ are indeed generated. These are quite laborious to keep track of, as they arise from multiple terms: the first, the third, the fourth, the sixth and the eighth in (\ref{eq:GravitinoEOMAnsatz}). Moreover, the first term alone produces several contributions coming from the various trombone-dependences of the covariant derivative of the spinor parameter, 
{\setlength\arraycolsep{2pt}
\begin{eqnarray} \label{eq:DerSusyParam}
D_\mu \epsilon^i &= & \partial_\mu \epsilon^i - \cV_{jk}{}^M \partial_\mu \cV_M{}^{ik} \epsilon^j + A_\mu^M \Big( \tfrac13  \cV_{jk}{}^N  \Theta_{MN}{}^P \cV_P{}^{ik} + \tfrac34 \cV_M{}^{ik} B_{jk} -\tfrac34 \cV_{Mjk} B^{ik}  \Big) \epsilon^j \nonumber \\[5pt]
&& -\tfrac14 \big( \omega_\mu{}^{ab} + 2  \, e_\mu{}^a e^{b \nu} A_\nu^M \vartheta_M   \big) \gamma_{ab} \, \epsilon^i -\tfrac12  A_\mu^M \vartheta_M \epsilon^i \; .
\end{eqnarray}
}Here, the fourth, fifth, seventh and eighth terms depend on $\vartheta_M$. After a long calculation, the relevant terms generated upon variation of (\ref{eq:GravitinoEOMAnsatz}) under (\ref{eq:susy-trans-fermions}), (\ref{eq:susy-trans-fermionsAnsatz}) add up altogether  to the l.h.s.~of 
{\setlength\arraycolsep{0pt}
\begin{eqnarray} \label{eq:BHeps}
&& \tfrac12\Big[ \big( -\tfrac13 \beta_2 + \tfrac{1}{24} \zeta + \tfrac12 \lambda -\tfrac34 \big) \, B^{ik} \cV_{Mjk} \epsilon^j 
+ \big( -\tfrac19 \beta_2 + \tfrac{1}{8} \zeta + 2\beta_1 + \tfrac34 \big) \, B_{jk} \cV_{M}{}^{ik} \epsilon^j \nonumber \\[4pt]
&& \quad + \big( \tfrac16 \beta_2 - \tfrac{1}{16} \zeta - \tfrac32 \big) \, B_{k\ell} \cV_{M}{}^{k\ell} \epsilon^i  \Big] H_{\nu\rho}^M \, \gamma^{\mu\nu\rho} = 0 \; .
\end{eqnarray}
}This must vanish as indicated by the reasons explained above.

The last type of terms that need to be tracked down involve contractions of ${\cal P}_\mu^{ijk\ell}$ and $B^{ij}$. Contributions of the schematic form $\gamma^{\mu\nu} \, {\cal P}_\nu  B \epsilon$ are generated by supersymmetry variation of the first, second and eighth terms in the gravitino equation (\ref{eq:GravitinoEOMAnsatz}), and must be required to vanish by the argument above. More precisely, one must enforce
\begin{equation} \label{eq:BPTermsGravitino}
i  \big( 12  \beta_1 -\tfrac43 \beta_2 +\tfrac12 \zeta  \big) \gamma^{\mu\nu} {\cal P}_\nu{}^{ijk\ell} B_{jk} \Omega_{\ell m} \epsilon^m = 0 \;  .
\end{equation}
In order to derive this expression, the following flow equation for $B^{ij}$ is needed,
\begin{equation}
D_\mu B^{ij} = -{\cal P}_\mu{}^{ijk\ell} B_{k \ell} \; ,
\end{equation}
which complements the analogue relations, (4.27) of \cite{deWit:2004nw}, for the non-trombone fermion shifts. Similar ${\cal P} B \epsilon$ terms arise from variation of the spin-$1/2$ equation (\ref{eq:FermionEOMAnsatz}). These do contribute to the vector equation of motion, and further relations for the coefficients are obtained from the requirement that they occur in the correct representation of USp$(8)$, the $\overline{\bm{27}}$. For this and other similar calculations, the explicit symplectic antisymmetrisations given in appendix \ref{sec:USp8Relations} come in handy. Specifically, some work shows that the first, second and ninth terms of (\ref{eq:FermionEOMAnsatz}) vary under (\ref{eq:susy-trans-fermions}), (\ref{eq:susy-trans-fermionsAnsatz}) as
{\setlength\arraycolsep{2pt}
\begin{eqnarray}
 \label{eq:FermionEOMVariation}  &&   \slashed{D}  \, \delta \chi^{ijk}  
  -\tfrac{i}{2} {\cal P}_\mu{}^{ijkl} \, \gamma^\nu \gamma^\mu \, \delta \psi_\nu{}^m \,\Omega_{lm} 
 -\tfrac12 \, \nu \,  i  B^{m[[i} \, \delta \chi^{jk]] n } \Omega_{mn}   
 + \tfrac{5}{3}   \tau \gamma^\mu \, \Omega^{[ij} {\cal P}_\mu^{k\ell mn]} B_{\ell m} \Omega_{np} \epsilon^p  \nonumber \\[5pt]
& =& -\tfrac{1}{12}  \,  \gamma^\mu \Big[ (12  \beta_2 -3\tau) \,  {\cal P}_\mu{}^{\ell m [ij} B_{\ell m} \epsilon^{k]}
+ (\nu - 4  \beta_2 +3\tau \big) \Omega^{[ij} {\cal P}_\mu{}^{k]\ell mn} B_{\ell m} \Omega_{np} \epsilon^p \nonumber \\
&& \qquad \quad - 3 (\nu +2\tau)  \,  \Omega^{\ell[i} {\cal P}_\mu{}^{jk]mn} B_{\ell m } \Omega_{np} \epsilon^p 
+ 2 (18 \beta_1 - \tau) \, {\cal P}_\mu{}^{ijk\ell} B_{\ell m} \epsilon^m \big] \; .
\end{eqnarray}
}The identically zero term $\Omega^{[ij} {\cal P}_\mu^{k\ell mn]} B_{\ell m} \Omega_{np} \epsilon^p \equiv 0$ (see (4.8) of \cite{deWit:2004nw}) has been added for convenience to both sides of (\ref{eq:FermionEOMVariation}), expanded out in manifestly antisymmetric pieces in $[ijk]$ on the r.h.s.,~and affected by an additional auxiliary coefficient $\tau$. Now, 
(\ref{eq:FermionEOMVariation}) must contribute to the vector equation of motion as 
$-\tfrac{1}{12}  \,  \gamma^\mu (12  \beta_2 -3\tau) \,  {\cal P}_\mu{}^{\ell m [[ij} B_{\ell m} \epsilon^{k]]}$, 
a requirement that leads to the following relations by comparison with (\ref{eq:STSample}):
\begin{equation} \label{eq:RelPBH}
12  \beta_2 -3\tau  = -3(\nu - 4  \beta_2 +3\tau ) \; ,\qquad
\nu +2\tau = 0 \; ,\qquad 
18 \beta_1 - \tau =0 \; .
\end{equation}
The system of linear equations for the coefficients provided by (\ref{eq:RelPBH}) together with the relations that can be read off from (\ref{eq:BDeps}), (\ref{eq:BHeps}) and (\ref{eq:BPTermsGravitino}) is overdetermined and, reassuringly, has a solution. It yields $\beta_1 = 1 $, $\beta_2= \tfrac98 $, confirming the result of section \ref{sec:susy}, together with $\lambda = 4 $, $\mu = \tfrac32 $, $\zeta = -21 $, $\nu = -36 $, along with $\tau = -36$ for the auxiliary parameter. These are the values for the physical parameters that have been brought to the main text.

%%%%%%%%%%%%%%%
%%%%%%%%%%%%%%%

\bibliography{references}
%%%%%%%%%%%%%%%
%%%%%%%%%%%%%%%

\end{document}